\documentclass[journal]{IEEEtran}
\usepackage{amsmath,amssymb,amsfonts,amsthm}
\usepackage[square, comma, sort&compress, numbers]{natbib}
\usepackage{algorithm, algpseudocode}
\usepackage{array}
\usepackage{multicol}
\usepackage{multirow,booktabs}
\usepackage{graphicx}
\usepackage{graphicx,pgf,tikz}
\usetikzlibrary{arrows,automata}

\ifCLASSOPTIONcompsoc
\usepackage[caption=false,font=normalsize,labelfont=sf,textfont=sf]{subfig}
\else
\usepackage[caption=false,font=footnotesize]{subfig}
\fi

\IEEEoverridecommandlockouts
\theoremstyle{definition}
\newtheorem{theorem}{Theorem}
\newtheorem{lemma}{Lemma}

\newtheorem{proposition}{Proposition}
\allowdisplaybreaks

\begin{document}
\title{Delay-Optimal Probabilistic Scheduling with \\ Arbitrary Arrival and Adaptive Transmission}

\author{
Xiang~Chen,~\IEEEmembership{Student~Member,~IEEE,}
Wei~Chen,~\IEEEmembership{Senior~Member,~IEEE,}
Joohyun~Lee,~\IEEEmembership{Member,~IEEE,}
and~Ness~B.~Shroff,~\IEEEmembership{Fellow,~IEEE}
\thanks{
X. Chen and W. Chen are with the Department of Electronic Engineering and Tsinghua National Laboratory for Information Science and Technology (TNList), Tsinghua University. E-mail: chen-xiang12@mails.tsinghua.edu.cn, wchen@tsinghua.edu.cn.

J. Lee is with the Department of ECE at The Ohio State
University. E-mail: lee.7119@osu.edu.

Ness B. Shroff holds a joint appointment in both the Department of ECE and the Department of CSE at The Ohio State University. E-mail: shroff@ece.osu.edu.
}}

\maketitle

\begin{abstract}
In this paper, we aim to obtain the optimal delay-power tradeoff and the corresponding optimal scheduling policy for arbitrary i.i.d. arrival process and adaptive transmissions. The number of backlogged packets at the transmitter is known to a scheduler, who has to determine how many backlogged packets to transmit during each time slot. The power consumption is assumed to be convex in transmission rates. Hence, if the scheduler transmits faster, the delay will be reduced but with higher power consumption. To obtain the optimal delay-power tradeoff and the corresponding optimal policy, we model the problem as a Constrained Markov Decision Process (CMDP), where we minimize the average delay given an average power constraint. By steady-state analysis and Lagrangian relaxation, we can show that the optimal tradeoff curve is decreasing, convex, and piecewise linear, and the optimal policy is threshold-based. Based on the revealed properties of the optimal policy, we develop an algorithm to efficiently obtain the optimal tradeoff curve and the optimal policy. The complexity of our proposed algorithm is much lower than a general algorithm based on Linear Programming. We validate the derived results and the proposed algorithm through Linear Programming and simulations.
\end{abstract}

\section{Introduction}
In this paper, we study an important problem of how to schedule the number of packets to transmit over a link taking into account both the delay and the power cost. This is an important problem because delay is a vital metric for many emerging applications (e.g., instant messenger, social network service, streaming media, and so on), and power consumption is critical to battery life of various mobile devices. In other words, we are studying the tradeoff between the timeliness and greenness of the communication service.

Such a delay-power scheduling problem can be formulated using a Markov Decision Process (MDP). The authors in \cite{collins1999transmission} were among the earliest who studied this type of scheduling problem. Specifically, they considered a two-state channel and finite time horizon. The dual problem was solved based on results derived by Dynamic Programming and induction. Follow-up papers \cite{berry2002communication, goyal2003power, bettesh2006optimal, berry2013optimal} extended this study in various directions. The optimal delay-power tradeoff curve is proven to be nonincreasing and convex in \cite{berry2002communication}. The existence of stationary optimal policy and the structure of the optimal policy are further investigated in \cite{goyal2003power}. Different types of power/rate control policies are studied in \cite{bettesh2006optimal}. In \cite{berry2013optimal}, the asymptotic small-delay regime is investigated. In \cite{rajan2004delay}, a piecewise linear delay-power tradeoff curve was obtained along with an approximate closed form expression.

If one can show monotonicity or a threshold type of structure to the optimal policy for MDPs, it helps to substantially reduce the computation complexity in finding the optimal policy. Indeed, the optimal scheduling policies are shown to be threshold-based or monotone in \cite{collins1999transmission, goyal2003power, berry2013optimal, agarwal2008structural, djonin2007mimo, ngo2010monotonicity, ata2005dynamic}, proven by studying the convexity, superadditivity / subadditivity, or supermodularity / submodularity of expected cost functions by induction using dynamic programming. However, most of these results are limited to the unconstrained Lagrangian Relaxation problem. In \cite{goyal2003power,ata2005dynamic}, some properties of the optimal policy for the constrained problem are described based on the results for the unconstrained problem. Detailed analysis on the optimal policy for the constrained problem is conducted in \cite{djonin2007mimo,ngo2010monotonicity}. In \cite{djonin2007mimo}, properties such as unichain policies and multimodularity of costs are assumed to be true so that monotone optimal policies can be proven. In \cite{ngo2010monotonicity}, the transmission action is either 1 or 0, i.e. to transmit or not. In order to obtain the detailed structure of the solution to the constrained problem, we believe that the analysis of the Lagrangian relaxation problem and the analysis of the structure of the delay-power tradeoff curve should be combined together.

In \cite{chen2015adaptive}, we study the optimal delay-power tradeoff problem. In particular, we minimize the average delay given an average power constraint, considering Bernoulli arrivals and adaptive transmissions. Some technical details are given in \cite{chen2016delay}, where we proved that the optimal tradeoff curve is convex and piecewise linear, and the optimal policies are threshold-based, by Constrained Markov Decision Process formulation and steady-state analysis. In this paper, we substantially generalize the Bernoulli arrival process to an arbitrary i.i.d. distribution. We show that the optimal policies for this generalized model are still threshold-based. Furthermore, we develop an efficient algorithm to find the optimal policy and the optimal delay-power tradeoff curve.

The remainder of this paper is organized as follows. The system model and the constrained problem are introduced in Section II. We show that the optimal policy is threshold-based in Section III by using steady-state analysis and Lagrangian relaxation. Based on theoretical results, we propose an efficient algorithm in Section IV to obtain the optimal tradeoff curve and the corresponding policies. In Section V, theoretical results and the proposed algorithm are verified by simulations. Section VI concludes the paper.

\section{System Model}
\begin{figure}[t]
\centering
\includegraphics[width=1\columnwidth]{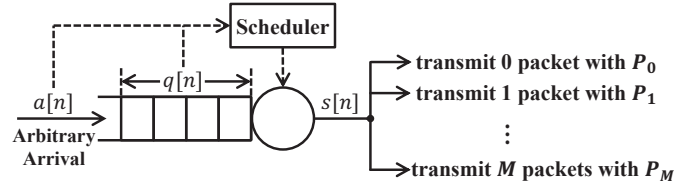}
\caption{System Model}
\label{fig_model}
\end{figure}

The system model is shown in \figurename~\ref{fig_model}. We assume there are $a[n]$ data packet(s) arriving at the end of the $n$th timeslot. The number $a[n]$ is i.i.d. for different values of $n$ and its distribution is given by $\text{Pr}\{a[n]=a\}=\alpha_a$, where $\alpha_a\ge 0$, $a\in\{0,1,\cdots,A\}$, and $\sum_{a=0}^A \alpha_a=1$. Therefore the expected number of packets arrived in each timeslot $n$ is given by $E_a= \sum_{a=0}^A a\alpha_a$.

Let $s[n]$ denote the number of data packets transmitted in timeslot $n$. Assume that at most $S$ packets can be transmitted in each timeslot because of the constraints of the transmitter, and $S \ge A$. Let $\tau[n]$ denote the transmission power consumed in timeslot $n$. Assume transmitting $s$ packet(s) will cost power $P_s$, where $s\in\{0,1,\cdots,S\}$, therefore $\tau[n]=P_{s[n]}$. Transmitting $0$ packet will cost no power, hence $P_0=0$. In typical communications, the power efficiency decreases as the transmission rate increases, hence we assume that $P_s$ is convex in $s$. Detailed explanations can be found in the Introduction section in \cite{chen2016delay}. The convexity of the power consumption function will be utilized in Theorem \ref{theorem_01} to prove that the optimal policy for the unconstrained problem is threshold-based.

Backlog packets are stored in a buffer with size $Q$. Let $q[n]\in\{0,1,\cdots,Q\}$ denote the queue length at the beginning of timeslot $n$. Since data arrive at the end of the timeslot, in order to avoid buffer overflow (i.e. $q[n]>Q$) and underflow (i.e. $q[n]<0$), we should have $0\le q[n]-s[n] \le Q-A$. Therefore the dynamics of the buffer is given as
\begin{align}
q[n+1]=q[n]-s[n]+a[n].
\end{align}

In timeslot $n$, we can decide how many packets to be transmitted based on the buffer state $q[n]$. It can be seen that this is a Markov Decision Process (MDP), where the queue length $q[n]$ is the state of the MDP, and the number of packets transmitted in each timeslot $s[n]$ is the action we take in each timeslot $n$. The probability distribution of the next state $q[n+1]$ is given by
\begin{align}
& \text{Pr}\{q[n+1]=j|q[n]=q,s[n]=s\}\nonumber\\
= &
\begin{cases}
\alpha_{j-q+s} & 0 \le j-q+s \le A,\\
0 & \text{otherwise}.
\end{cases}
\end{align}

We minimize the average queueing delay given an average power constraint, which makes it a Constrained Markov Decision Process (CMDP). For an infinite-horizon CMDP with stationary parameters, according to \cite[Theorem 11.3]{altman1999constrained}, stationary policies are complete, which means stationary policies can achieve the optimal performance. Therefore we only need to consider stationary policies in this problem. Let $f_{q,s}$ denote the probability to transmit $s$ packet(s) when $q[n]=q$, i.e.,
\begin{align}
f_{q,s}=\text{Pr}\{s[n]=s|q[n]=q\}.
\end{align}
Then we have $\sum_{s=0}^{S}f_{q,s}=1$ for $q=0,\cdots,Q$. Since we guarantee that the transmission strategy will avoid overflow or underflow, we set
\begin{align}
f_{q,s}=0 \text{ if } q-s<0 \text{ or } q-s>Q-A.
\label{eq_f0}
\end{align}

Let $\boldsymbol{F}$ denote a $(Q+1)\times(S+1)$ matrix whose element in the $(q+1)$th row and the $(s+1)$th column is $f_{q,s}$. Therefore matrix $\boldsymbol{F}$ can represent a stationary transmission policy. Let $P_{\boldsymbol{F}}$ and $D_{\boldsymbol{F}}$ denote the average power consumption and the average queueing delay under policy $\boldsymbol{F}$. Let $\mathcal{F}$ denote the set of all feasible stationary policies that guarantee no queue overflow or underflow. Let $\mathcal{F}_D$ denote the set of all stationary and deterministic policies which can guarantee no overflow or underflow. Thus to obtain the optimal tradeoff curve, we can minimize the average delay given an average power constraint $P_{\text{th}}$ shown as
\begin{subequations}
\label{eqn_stationary_optimization}
\begin{align}
\min\limits_{\boldsymbol{F}\in\mathcal{F}}\quad & D_{\boldsymbol{F}}\\
\text{s.t.}\quad & P_{\boldsymbol{F}} \le P_{\text{th}}.
\end{align}
\end{subequations}

From another perspective, policy $\boldsymbol{F}$ will determine a point $Z_{\boldsymbol{F}}=(P_{\boldsymbol{F}},D_{\boldsymbol{F}})$ in the delay-power plane. Define $\mathcal{R}=\{Z_{\boldsymbol{F}} | \boldsymbol{F} \in \mathcal{F} \}$ as the set of all feasible points in the delay-power plane. Intuitively, since the power consumption for each data packet increases if we want to transmit faster, there is a tradeoff between the average queueing delay and the average power consumption. Thus the optimal delay-power tradeoff curve can be presented as $\mathcal{L}=\{(P,D)\in\mathcal{R}|\forall(P',D')\in\mathcal{R},\text{ either }P'\ge P\text{ or }D'\ge D\}$.

\begin{figure*}[!t]
\centering
\begin{tikzpicture}[->, >=stealth', very thick, every node/.style={fill=white, font=\normalsize}]
\node[state] (0) at (0,0) {$0$};
\node[state] (1) at (3.2,0) {$1$};
\node[state] (2) at (6.4,0) {$2$};
\node[state] (3) at (9.6,0) {$3$};
\node[state] (4) at (12.8,0) {$4$};
\node[state] (5) at (16,0) {$5$};

\path
(1) edge [bend right=15] node {$\lambda_{1,0}$} (0)
(2) edge [bend right=15] node {$\lambda_{2,1}$} (1)
(3) edge [bend right=15] node {$\lambda_{3,2}$} (2)
(4) edge [bend right=15] node {$\lambda_{4,3}$} (3)
(5) edge [bend right=15] node {$\lambda_{5,4}$} (4)

(2) edge [bend right=28] node {$\lambda_{2,0}$} (0)
(3) edge [bend right=28] node {$\lambda_{3,1}$} (1)
(4) edge [bend right=28] node {$\lambda_{4,2}$} (2)
(5) edge [bend right=28] node {$\lambda_{5,3}$} (3)

(3) edge [bend right=40] node {$\lambda_{3,0}$} (0)
(4) edge [bend right=40] node {$\lambda_{4,1}$} (1)
(5) edge [bend right=40] node {$\lambda_{5,2}$} (2)

(0) edge [bend right=15] node {$\lambda_{0,1}$} (1)
(1) edge [bend right=15] node {$\lambda_{1,2}$} (2)
(2) edge [bend right=15] node {$\lambda_{2,3}$} (3)
(3) edge [bend right=15] node {$\lambda_{3,4}$} (4)
(4) edge [bend right=15] node {$\lambda_{4,5}$} (5)

(0) edge [bend right=28] node {$\lambda_{0,2}$} (2)
(1) edge [bend right=28] node {$\lambda_{1,3}$} (3)
(2) edge [bend right=28] node {$\lambda_{2,4}$} (4)
(3) edge [bend right=28] node {$\lambda_{3,5}$} (5)

(0) edge [bend right=40] node {$\lambda_{0,3}$} (3)
(1) edge [bend right=40] node {$\lambda_{1,4}$} (4)
(2) edge [bend right=40] node {$\lambda_{2,5}$} (5)
;
\end{tikzpicture}
\caption{Markov Chain of $t[n]$ ($Q=5$, $A=3$, $M=3$, $\lambda_{i,i}$ for all $i$ are omitted to keep the diagram legible)}
\label{fig_markov}
\end{figure*}
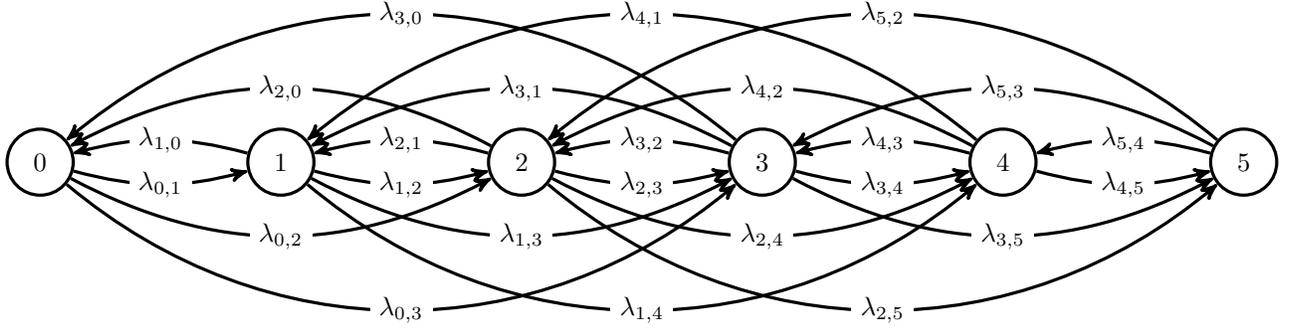

If we fix a stationary policy for a Markov Decision Process, the Markov Decision Process will degenerate to a Markov Reward Process (MRP). Let $\lambda_{i,j}$ denote the transition probability from state $i$ to state $j$. According to the system model, because of the constraints of transmission and arrival processes, the state transition probability can be derived as
\begin{align}
\lambda_{i,j}=\sum_{s=\max\{0,i+A-Q,i-j\}}^{\min\{S,i,i-j+A\}}\alpha_{j-i+s}f_{i,s}.
\end{align}
An example of the transition diagram is shown in \figurename~\ref{fig_markov}, where $\lambda_{i,i}$ for $i=0,\cdots,Q$ are omitted to keep the diagram legible.

The Markov chain could have more than one closed communication classes under certain transmission policies. Under this circumstance, the limiting probability distribution and the average cost are dependent on the initial state and the sample paths. In Appendix \ref{appen_unichain}, it is proven that we only need to consider the cases where the Markov chain has only one closed communication class, which is called a unichain. Becausae of this key result, we focus only on the unichain cases in the following.

\section{Optimal Threshold-Based Policy for the Constrained Markov Decision Process}

In this section, we will demonstrate that the optimal policy for the Constrained MDP problem is threshold-based. In other words, for an optimal policy, more data will be transmitted if the queue is longer. We give the rigorous definition of a stationary threshold-based policy $\boldsymbol{F}$ that, there exist $(S+1)$ thresholds $0 \le q_{\boldsymbol{F}}(0) \le q_{\boldsymbol{F}}(1) \le \cdots \le q_{\boldsymbol{F}}(S) \le Q$, such that $f_{q,s}>0$ only when $q_{\boldsymbol{F}}(s-1) \le q \le q_{\boldsymbol{F}}(s)$ (set $q_{\boldsymbol{F}}(-1)=-1$ for simplicity of notation). According to this definition, under policy $\boldsymbol{F}$, when the queue state is larger than threshold $q_{\boldsymbol{F}}(s-1)$ and smaller than $q_{\boldsymbol{F}}(s)$, it transmits $s$ packet(s). When the queue state is equal to threshold $q_{\boldsymbol{F}}(s)$, it transmits $s$ or $(s+1)$ packet(s). Note that under this definition, probabilistic policies can also be threshold-based.

In the following, we will first conduct the steady-state analysis of the Markov process, based on which we can show the properties of the feasible delay-power region and the optimal delay-power tradeoff, and then by proving that the Lagrangian relaxation problem has a deterministic and threshold-based optimal policy, we can finally show that the optimal policy for the constrained problem is threshold-based.

\subsection{Steady State Analysis}
Since we can focus on unichain cases, which contain a single recurrent class plus possibly some transient states, the steady-state probability distribution exists for the Markov process. Let $\pi_{\boldsymbol{F}}(q)$ denote the steady-state probability for state $q$ when applying policy $\boldsymbol{F}$. Set $\boldsymbol{\pi}_{\boldsymbol{F}}=[\pi_{\boldsymbol{F}}(0),\cdots,\pi_{\boldsymbol{F}}(Q)]^T$. Define $\boldsymbol{\Lambda}_{\boldsymbol{F}}$ as a $(Q+1)\times(Q+1)$ matrix whose element in the $(i+1)$th column and the $(j+1)$th row is $\lambda_{i,j}$, which is determined by policy $\boldsymbol{F}$. Set $\boldsymbol{I}$ as the identity matrix. Define $\boldsymbol{1}=[1,\cdots,1]^T$, and $\boldsymbol{0}=[0,\cdots,0]^T$. Set $\boldsymbol{G}_{\boldsymbol{F}}=\boldsymbol{\Lambda}_{\boldsymbol{F}}-\boldsymbol{I}$. Set
$\boldsymbol{H}_{\boldsymbol{F}}=\left[
\begin{array}{c}
\boldsymbol{1}^T\\
\boldsymbol{G}_{\boldsymbol{F}}(0:(Q-1),:)
\end{array}
\right]
$ and
$\boldsymbol{c}=\left[
\begin{array}{c}
1\\\boldsymbol{0}
\end{array}
\right]$.

According to the definition of the steady-state distribution, we have $\boldsymbol{G}_{\boldsymbol{F}}\boldsymbol{\pi}_{\boldsymbol{F}}=\boldsymbol{0}$ and $\boldsymbol{1}^T\boldsymbol{\pi}_{\boldsymbol{F}}=1$. For a unichain, the rank of $\boldsymbol{G}_{\boldsymbol{F}}$ is $Q$. Therefore, we have $\boldsymbol{H}_{\boldsymbol{F}}$ is invertible and
\begin{align}
\boldsymbol{H}_{\boldsymbol{F}}\boldsymbol{\pi}_{\boldsymbol{F}}=\boldsymbol{c}.
\label{pi-H}
\end{align}

For state $q$, transmitting $s$ packet(s) will cost $P_s$ with probability $f_{q,s}$. Define $\boldsymbol{p}_{\boldsymbol{F}}=[\sum_{s=0}^S P_s f_{0,s},\cdots,\sum_{s=0}^S P_s f_{Q,s}]^T$, which is a function of $\boldsymbol{F}$. The average power consumption $P_{\boldsymbol{F}}$ can be expressed as
\begin{align}
P_{\boldsymbol{F}}=\sum_{q=0}^{Q} \pi_{\boldsymbol{F}}(q) \sum_{s=0}^S P_s f_{q,s}=\boldsymbol{p}_{\boldsymbol{F}}^T\boldsymbol{\pi}_{\boldsymbol{F}}.
\label{PwithPi}
\end{align}
Similarly, define $\boldsymbol{d}=[0,1,\cdots,Q]^T$. According to Little's Law, the average delay $D_{\boldsymbol{F}}$ under policy $\boldsymbol{F}$ is
\begin{align}
D_{\boldsymbol{F}}=\frac{1}{E_a}\sum_{q=0}^{Q} q \pi_{\boldsymbol{F}}(q)=\frac{1}{E_a}\boldsymbol{d}^T\boldsymbol{\pi}_{\boldsymbol{F}}.
\label{DwithPi}
\end{align}

The following theorem describes the structure of the feasible delay-power region and the optimal delay-power tradeoff curve.

\begin{theorem}
The set of all feasible points in the delay-power plane, $\mathcal{R}$, and the optimal delay-power tradeoff curve $\mathcal{L}$, satisfy that
\begin{enumerate}
\item The set $\mathcal{R}$ is a convex polygon.
\item The curve $\mathcal{L}$ is piecewise linear, decreasing, and convex.
\item Vertices of $\mathcal{R}$ and $\mathcal{L}$ are all obtained by deterministic scheduling policies.
\item The policies corresponding to adjacent vertices of $\mathcal{R}$ and $\mathcal{L}$ take different actions in only one state.
\end{enumerate}
\label{theorem_piecewise_linear}
\end{theorem}
\begin{IEEEproof}
See Appendix \ref{appen_piecewiselinear}.
\end{IEEEproof}

\subsection{Optimal Deterministic Threshold-Based Policy for the Lagrangian Relaxation Problem}

In (\ref{eqn_stationary_optimization}), we formulate the optimization problem as a Constrained MDP, which is difficult to solve in general. Let $\mu\ge 0$ denote the Lagrange multiplier. Consider the Lagrangian relaxation of (\ref{eqn_stationary_optimization})
\begin{align}
\min\limits_{\boldsymbol{F}\in \mathcal{F}} \quad D_{\boldsymbol{F}}+\mu P_{\boldsymbol{F}}-\mu P_{\text{th}}.
\label{eqn_lag_relax}
\end{align}

In (\ref{eqn_lag_relax}), the term $-\mu P_{\text{th}}$ is constant. Therefore, the Lagrangian relaxation problem is minimizing the weighted average cost $D_{\boldsymbol{F}}+\mu P_{\boldsymbol{F}}$, which becomes an unconstrained infinite-horizon Markov Decision Process with an average cost criterion. It is proven in \cite[Theorem 9.1.8]{puterman2014markov} that, there exists an optimal stationary deterministic policy. Moreover, the optimal policy for the relaxation problem has the following property.

\begin{theorem}
An optimal policy $\boldsymbol{F}$ for the unconstrained Markov Decision Process is threshold-based. That is to say, there exists $(S+1)$ thresholds $q_{\boldsymbol{F}}(0) \le q_{\boldsymbol{F}}(1) \le \cdots \le q_{\boldsymbol{F}}(S)$, such that
\begin{align}
\begin{cases}
f_{q,s}=1 & q_{\boldsymbol{F}}(s-1)<q\le q_{\boldsymbol{F}}(s), s=0,\cdots,S\\
f_{q,s}=0 & \text{otherwise}
\end{cases}
\label{eqn_deterministic_threshold}
\end{align}
where $q_{\boldsymbol{F}}(-1)=-1$.
\label{theorem_01}
\end{theorem}
\begin{IEEEproof}
See Appendix \ref{appen_unconstrained_threshold}.
\end{IEEEproof}

\subsection{Optimal Threshold-Based Policy for the Constrained Problem}
\begin{figure}[t]
\centering
\includegraphics[width=0.9\columnwidth]{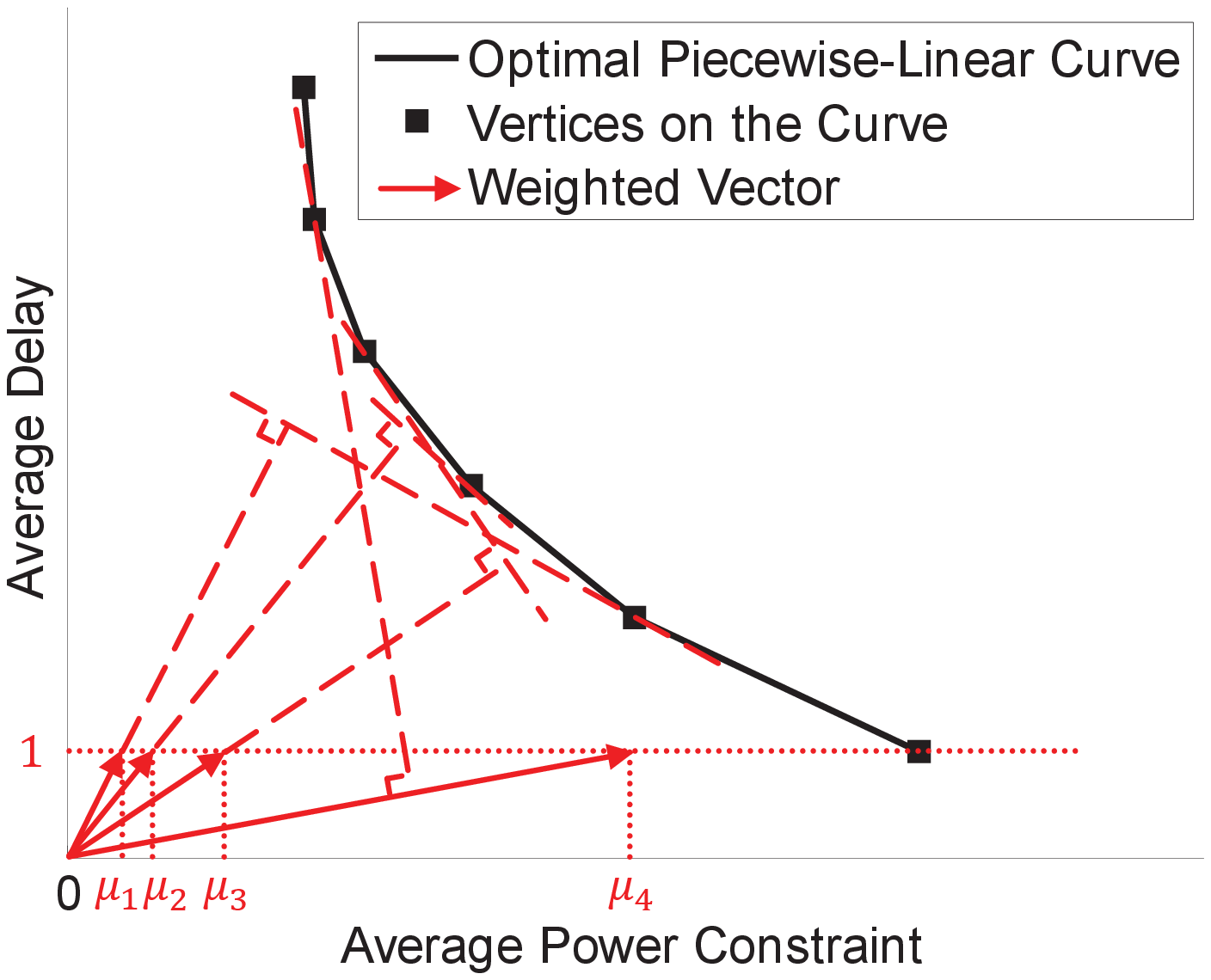}
\caption{The minimum inner product of points on $\mathcal{L}$ and the weighted vector can always be obtained by vertices of $\mathcal{L}$}
\label{fig_weightedsum_piecewiselinear}
\end{figure}

From another perspective, $D_{\boldsymbol{F}}+\mu P_{\boldsymbol{F}}=\langle(\mu, 1),(P_{\boldsymbol{F}},D_{\boldsymbol{F}})\rangle$ can be seen as the inner product of vector $(\mu, 1)$ and $Z_{\boldsymbol{F}}$. Since $\mathcal{L}$ is piecewise linear, decreasing and convex, the corresponding $Z_{\boldsymbol{F}}$ minimizing the inner product will be obtained by the vertices of $\mathcal{L}$, as can be observed in \figurename~\ref{fig_weightedsum_piecewiselinear}. Since the conclusion in Theorem \ref{theorem_01} holds for any $\mu$, the vertices of the optimal tradeoff curve can all be obtained by optimal policies for the Lagrangian relaxation problem, which are deterministic and threshold-based. Moreover, from Theorem \ref{theorem_piecewise_linear}, the adjacent vertices of $\mathcal{L}$ are obtained by policies which take different actions in only one state. Therefore, we can have the following theorem.

\begin{theorem}
Given an average power constraint, the scheduling policy $\boldsymbol{F}$ to minimize the average delay takes the following form: there exists $(S+1)$ thresholds $q_{\boldsymbol{F}}(0) \le q_{\boldsymbol{F}}(1) \le \cdots \le q_{\boldsymbol{F}}(S)$, one of which we name $q_{\boldsymbol{F}}(s^*)$, such that
\begin{align}
\left\{
\begin{array}{@{}ll}
f_{q,s}=1 & q_{\boldsymbol{F}}(s-1)<q\le q_{\boldsymbol{F}}(s),s\neq s^*\\
f_{q,s^*}=1\qquad\qquad & q_{\boldsymbol{F}}(s^*-1)<q< q_{\boldsymbol{F}}(s^*)\\
\multicolumn{2}{@{}l}{f_{q_{\boldsymbol{F}}(s^*),s^*}+f_{q_{\boldsymbol{F}}(s^*),s^*+1}=1} \\
f_{q,s}=0 & \text{otherwise}
\end{array}
\right.
\label{eqn_threshold}
\end{align}
where $q_{\boldsymbol{F}}(-1)=-1$.
\label{theorem_threshold}
\end{theorem}
\begin{IEEEproof}
Since the optimal tradeoff curve is piecewise linear, assume $Z_{\boldsymbol{F}}$ is on the line segment between vertices $Z_{\boldsymbol{F}'}$ and $Z_{\boldsymbol{F}''}$. According to Theorem \ref{theorem_01}, the form of optimal policies $\boldsymbol{F}'$ and $\boldsymbol{F}''$, which are corresponding to vertices of the optimal tradeoff curve, satisfies (\ref{eqn_deterministic_threshold}). Moreover, according to Theorem \ref{theorem_piecewise_linear}, the policies corresponding to adjacent vertices of $\mathcal{L}$ take different actions in only one state. Define the thresholds for $\boldsymbol{F}'$ as $q_{\boldsymbol{F}'}(0),q_{\boldsymbol{F}'}(1),\cdots,q_{\boldsymbol{F}'}(s^*),\cdots,q_{\boldsymbol{F}'}(S)$, then the thresholds for $\boldsymbol{F}'$ can be expressed as $q_{\boldsymbol{F}'}(0),q_{\boldsymbol{F}'}(1),\cdots,q_{\boldsymbol{F}'}(s^*)-1,\cdots,q_{\boldsymbol{F}'}(S)$, where the two policies take different actions only in state $q_{\boldsymbol{F}'}(s^*)$. Since $Z_{\boldsymbol{F}}$, the policy to obtain a point on the line segment between $Z_{\boldsymbol{F}'}$ and $Z_{\boldsymbol{F}''}$ is the convex combination of $\boldsymbol{F}'$ and $\boldsymbol{F}''$, it should have the form shown in (\ref{eqn_threshold}).
\end{IEEEproof}

We can see that the optimal policy for the Constrained Markov Decision Process may not be deterministic. At most two elements in the policy matrix $\boldsymbol{F}$, i.e. $f_{q_{\boldsymbol{F}}(s^*),s^*}$ and $f_{q_{\boldsymbol{F}}(s^*),s^*+1}$, can be decimal, while the other elements are either 0 or 1. Policies in this form also satisfy our definition of stationary threshold-based policy at the beginning of Section III.

\section{Algorithm to Efficiently Obtain the Optimal Tradeoff Curve}
We design Algorithm \ref{algo_curve} to efficiently obtain the optimal delay-power tradeoff curve and the corresponding optimal policies. Similar to \cite{chen2016delay}, this algorithm takes advantage of the properties we have shown, i.e., the optimal delay-power tradeoff curve is piecewise linear, the vertices are obtained by deterministic threshold-based policies, and policies corresponding to two adjacent vertices take different actions in only one state. Therefore given the optimal policy for a certain vertex, we can narrow down the alternatives of optimal policies for its adjacent vertex. The policies corresponding to points between two adjacent vertices can also be easily generated.

\begin{algorithm}[t]
\caption{Constructing the Optimal Delay-Power Tradeoff}
\begin{algorithmic}[1]
\State Construct $\boldsymbol{F}$ whose thresholds $q_{\boldsymbol{F}}(s)=s$ for $s < A$ and $q_{\boldsymbol{F}}(s)=Q$ for $s \ge A$
\State Calculate $D_{\boldsymbol{F}}$ and $P_{\boldsymbol{F}}$
\State $\mathcal{F}_c \gets [\boldsymbol{F}]$, $D_c \gets D_{\boldsymbol{F}}$, $P_c \gets D_{\boldsymbol{F}}$
\While{$\mathcal{F}_c \neq \emptyset$}
\State $\mathcal{F}_p \gets \mathcal{F}_c$, $D_p \gets D_c$, $P_p \gets D_c$
\State $\mathcal{F}_c \gets \emptyset$, $slope \gets +\infty$
\While{$\mathcal{F}_p \neq \emptyset$}
\State{$\boldsymbol{F}$=$\mathcal{F}_p$.pop(0)}
\ForAll{$0 < s^* < A$}
\State
\begin{tabular}{@{}l}
Construct $\boldsymbol{F}'$ where $q_{\boldsymbol{F}'}(s^*)=q_{\boldsymbol{F}}(s^*)+1$ \\ and $q_{\boldsymbol{F}'}(s)=q_{\boldsymbol{F}}(s)$ for $s\neq s^*$
\end{tabular}
\If{$\boldsymbol{F}'$ is feasible and threshold-based}
\State Calculate $D_{\boldsymbol{F}'}$ and $P_{\boldsymbol{F}'}$
\If{$D_{\boldsymbol{F}'} = D_p$ and $P_{\boldsymbol{F}'} = P_p$}
\State $\mathcal{F}_p$.append$(\boldsymbol{F}')$
\ElsIf{$D_{\boldsymbol{F}'} \ge D_p$ and $P_{\boldsymbol{F}'} < P_p$}
\If{$\frac{D_{\boldsymbol{F}'}-D_p}{P_p-P_{\boldsymbol{F}'}}<slope$}
\State $\mathcal{F}_c \gets [\boldsymbol{F}']$, $slope \gets \frac{D_{\boldsymbol{F}'}-D_p}{P_p-P_{\boldsymbol{F}'}}$
\State $D_c \gets D_{\boldsymbol{F}'}$, $P_c \gets P_{\boldsymbol{F}'}$
\ElsIf{$\frac{D_{\boldsymbol{F}'}-D_p}{P_p-P_{\boldsymbol{F}'}}=slope$}
\If{$P_{\boldsymbol{F}'}=P_c$}
\State $\mathcal{F}_c$.append$(\boldsymbol{F}')$
\ElsIf{$P_{\boldsymbol{F}'}>P_c$}
\State $\mathcal{F}_c \gets [\boldsymbol{F}']$, $slope \gets \frac{D_{\boldsymbol{F}'}-D_p}{P_p-P_{\boldsymbol{F}'}}$
\State $D_c \gets D_{\boldsymbol{F}'}$, $P_c \gets P_{\boldsymbol{F}'}$
\EndIf
\EndIf
\EndIf
\EndIf
\EndFor
\EndWhile
\State
\begin{tabular}{@{}l}
Draw the line segment connecting $(P_p,D_p)$ and \\ $(P_c,D_c)$
\end{tabular}
\EndWhile
\end{algorithmic}
\label{algo_curve}
\end{algorithm}

Our proposed iterative algorithm starts from the bottom-right vertex of the optimal tradeoff curve, whose corresponding policy is known to transmit as much as possible. Then for each vertex we have determined, we enumerate the candidates for the next vertex. According to the properties we have obtained, we only need to search for deterministic threshold-based policies which take different actions in only one threshold. By comparing all the candidates, the next vertex will be determined by the policy candidate whose connecting line with the current vertex has the minimum absolute slope and the minimum length. Note that a vertex can be obtained by more than one policy, therefore we use lists $\mathcal{F}_p$ and $\mathcal{F}_c$ to restore all policies corresponding to the previous and the current vertices.

The complexity of this algorithm is much smaller than using general methods. Since during each iteration, one of the thresholds of the optimal policy will be decreased by 1, the maximum iteration times are $AQ$. Within each iteration, we have $A$ thresholds to try. For each candidate, the most time consuming operation, i.e. the matrix inversion, costs $\boldsymbol{O}(Q^3)$. Therefore the complexity of the algorithm is $\boldsymbol{O}(A^2Q^4)$.

In comparison, we also formulate a Linear Programming (LP) to obtain the optimal tradeoff curve. As demonstrated in \cite[Chapter 11.5]{altman1999constrained}, all CMDP problems with infinite horizon and average cost can be formulated as Linear Programming. In our case, by taking $x_{q,s}=\pi(q) f_{q,s}$ as variables, we can formulate an LP with $QS$ variables to minimize the average delay given a certain power constraint. Due to space limitation, we provide the LP without explanations.
\begin{subequations}
\label{eqn_lp}
\begin{align}
\min\quad
& \frac{1}{E_a}\sum_{q=0}^{Q} q \sum_{s=0}^S x_{q,s}\\
\text{s.t.}\quad
& \sum_{q=0}^{Q} \sum_{s=0}^S P_s x_{q,s} \le P_{\text{th}}\\
& \sum_{l=\max\{0,q-A\}}^{q-1}\sum_{a=0}^{A}\sum_{s=0}^{l+a-q} \alpha_a x_{l,s} \nonumber\\
&=\sum_{r=q}^{\min\{q+S-1,Q\}}\sum_{a=0}^{A}\sum_{s=r+a-q+1}^{S} \alpha_a x_{r,s} \quad q=1,\cdots,Q\\
& \sum_{q=0}^{Q} \sum_{s=0}^S x_{q,s}=1\\
& x_{q,s}=0 \qquad \forall q-s<0 \text{ or } q-s>Q-A\\
& x_{q,s}\ge 0 \qquad \forall 0 \le q-s \le Q-A.
\end{align}
\end{subequations}

By solving the LP, we can obtain a point on the optimal tradeoff curve. If we apply the ellipsoid algorithm to solve the LP problem, the computational complexity is $\boldsymbol{O}(S^4Q^4)$. It means that, the computation to obtain one point on the optimal tradeoff curve by applying LP is larger than obtaining the entire curve with our proposed algorithm. This demonstrates the inherent advantage of using the revealed properties of the optimal tradeoff curve and the optimal policies.

\begin{figure}[t] 
\centering
\includegraphics[width=0.9\columnwidth]{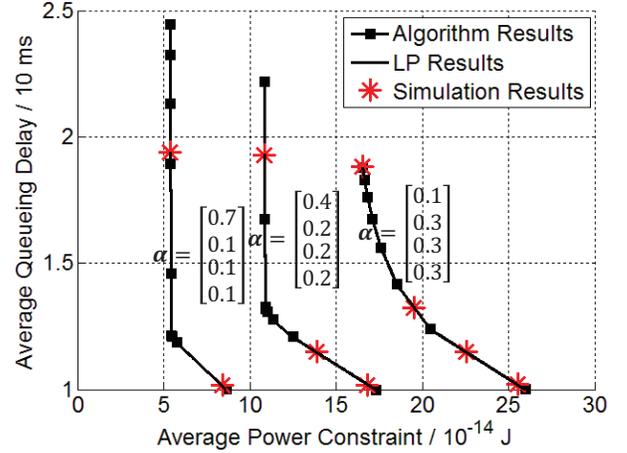}
\caption{Optimal Delay-Power Tradeoff Curves}
\label{fig_optimization_1}
\end{figure}

\begin{figure}[t] 
\centering
\includegraphics[width=0.9\columnwidth]{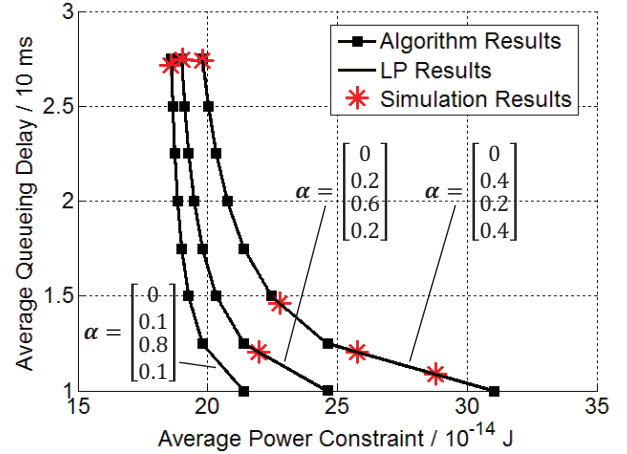}
\caption{Optimal Delay-Power Tradeoff Curves with the Same Arrival Rate}
\label{fig_optimization_2}
\end{figure}

\section{Numerical Results}
In this section, we validate our theoretical results and the proposed algorithm by conducting LP numerical computation and simulations. We consider a practical scenario with adaptive M-PSK transmissions. The optional modulations are BPSK, QPSK, and 8-PSK. Assume the bandwidth = 1 MHz, the length of a timeslot = 10 ms, and the target bit error rate ber=$10^{-5}$. Assume a data packet contains 10,000 bits, and in each timeslot the number of arriving packet could be 0, 1, 2 or 3. Then by adaptively applying BPSK, QPSK, or 8-PSK, we can respectively transmit 1, 2, or 3 packets in a timeslot, which means $S=3$. Assume the one-sided noise power spectral density $N_0$=-150 dBm/Hz. The transmission power for different transmission rates can be calculated as $P_0=0$ J, $P_1=9.0\times 10^{-14}$ J, $P_2=18.2\times 10^{-14}$ J, and $P_3=59.5\times 10^{-14}$ J. Set the buffer size as $Q=100$.

The optimal delay-power tradeoff curves are shown in \figurename~\ref{fig_optimization_1} and \figurename~\ref{fig_optimization_2}. In each figure, we vary the arrival process to get different tradeoff curves. As can be observed, the tradeoff curves generated by Algorithm \ref{algo_curve} perfectly match the Linear Programming and simulation results. As proven in Theorem \ref{theorem_piecewise_linear}, the optimal tradeoff curves are piecewise linear, decreasing, and convex. The vertices of the curves obtained by Algorithm \ref{algo_curve} are marked by squares. The corresponding optimal policies can be checked as threshold-based. The minimum average delay is 1 for all curves, because when we transmit as much as we can, all data packets will stay in the queue for exactly one timeslot. In \figurename~\ref{fig_optimization_1}, with the average arrival rate increasing, the curve gets higher because of the heavier workload. In \figurename~\ref{fig_optimization_2}, the three arrival processes have the same average arrival rate and different variance. When the variance gets larger, it is more likely that the queue size gets long in a short time duration, which leads to higher delay. It is interesting to characterize the effect of the variance in the arrival process, which we leave as a future work.

\section{Conclusion}
In this paper, we extend our previous work to obtain the optimal delay-power tradeoff and the corresponding optimal scheduling policy considering arbitrary i.i.d. arrival and adaptive transmissions. The scheduler optimize the transmission in each timeslot according to the buffer state. We formulate this problem as a CMDP, and minimize the average delay to obtain the optimal tradeoff curve. By studying the steady-state properties and the Lagrangian relaxation of the CMDP problem, we can prove that the optimal delay-power tradeoff curve is convex and piecewise linear, on which the adjacent vertices are obtained by policies taking different actions in only one state. Based on this, the optimal policies are proven to be threshold-based. We also design an efficient algorithm to obtain the optimal tradeoff curve and the optimal policies. Linear Programming and simulations are conducted to confirm the theoretical results and the proposed algorithm.

\appendices

\section{Proof of the Equivalency to Reduce to Unichain cases}
\label{appen_unichain}
We claim that we can focus only on the unichain cases, because for any Markov process with multiple recurrent classes determined by a certain policy, we can design a policy which leads to a unichain Markov process having the same performance as any of the recurrent class. We strictly express the reason as a proposition below, and give the detailed proof.

\begin{proposition}
In the Markov Decision Process with arbitrary arrival and adaptive transmission, if there is more than one closed communication class in the Markov chain generated by policy $\boldsymbol{F}$, which we define as $\mathcal{C}_1$, $\cdots$, $\mathcal{C}_L$ where $L>1$, then for any $1\le l \le L$, there exists a policy $\boldsymbol{F}_l$, under which the Markov chain has $\mathcal{C}_l$ as its only closed communication class. Furthermore, the steady-state distribution and the average cost of the Markov chain under $\boldsymbol{F}$ starting from state $c\in\mathcal{C}_l$ are the same as the steady-state distribution and the average cost of the Markov chain under $\boldsymbol{F}_l$.
\end{proposition}
\begin{IEEEproof}
Define the set of those transient states that have access to $\mathcal{C}_l$ as $\mathcal{C}_l^t$. Define the set of transient states which don't have access to $\mathcal{C}_l$ as $\mathcal{C}_{nl}^t$. Therefore $\{\mathcal{C}_1,\cdots,\mathcal{C}_L,\mathcal{C}_l^t,\mathcal{C}_{nl}^t\}$ is a partition of the states of the MDP. There should exists at least one state $c\in\bigcup_{i=1,i\neq l}^{+\infty} \mathcal{C}_i \cup \mathcal{C}_{nl}^t$ which is next to a state $c'\in \mathcal{C}_l \cup \mathcal{C}_l^t$. We can always change the action in state $c$ such that state $c$ can access the set $\mathcal{C}_l \cup \mathcal{C}_l^t$. After the modification, state $c$ will be a transient state which has access to $\mathcal{C}_l$. The states which communicate with $c$ will also be transient states which have access to $\mathcal{C}_l$.

We update the partition of states since the policy is changed. According to the above description, the set $\mathcal{C}_l$ won't change, while the cardinality of $\mathcal{C}_l^t$ will be strictly increasing. Hence, by repeating the above operation for finite times, every state of the MDP will be partitioned in either $\mathcal{C}_l$ or $\mathcal{C}_l^t$. The Markov chain generated by the modified policy has $\mathcal{C}_l$ as its only closed communication class, and the modified policy is the $\boldsymbol{F}_l$ we request.

Since the actions of states in $\mathcal{C}_l$ are the same for policy $\boldsymbol{F}$ and $\boldsymbol{F}_l$, the steady-state distribution and the average cost corresponding to policy $\boldsymbol{F}$ starting from state $c\in\mathcal{C}_l$ are the same as those under policy $\boldsymbol{F}_l$.
\end{IEEEproof}

\section{Proof of Theorem 1}
\label{appen_piecewiselinear}

In order to prove Theorem 1, we will first prove a lemma showing that the mapping from $\boldsymbol{F}$ to $Z_{\boldsymbol{F}}=(P_{\boldsymbol{F}},D_{\boldsymbol{F}})$ has a partially linear property in the first subsection. In the second subsection, we will prove that the set $\mathcal{R}$ is a convex polygon, whose vertices are all obtained by deterministic scheduling policies, and the policies corresponding to adjacent vertices of $\mathcal{R}$ take different actions in only one state. In the third subsection, we will prove that the set $\mathcal{L}$ is piecewise linear, decreasing, and convex, whose vertices are obtained by deterministic scheduling policies, and the policies corresponding to adjacent vertices of $\mathcal{L}$ take different actions in only one state.

In correspondence with Theorem 1, conclusion 1) in the theorem is proven in Subsection B, conclusion 2) is proven in Subsection C, and conclusion 3) and 4) are proven by combining results in Subsection B and C.

\subsection{Partially Linear Property of Scheduling Policies}

\begin{lemma}
$\boldsymbol{F}$ and $\boldsymbol{F}'$ are two policies different only when $q[n]=q$, i.e., these two matrices are different only in the $(q+1)$th row. Denote $\boldsymbol{F}''=(1-\epsilon)\boldsymbol{F}+\epsilon\boldsymbol{F}'$ where $0\le \epsilon\le 1$. Then\newline
1) There exists a certain $0\le \epsilon'\le 1$ so that $P_{\boldsymbol{F}''}=(1-\epsilon')P_{\boldsymbol{F}}+\epsilon' P_{\boldsymbol{F}'}$ and $D_{\boldsymbol{F}''}=(1-\epsilon')D_{\boldsymbol{F}}+\epsilon' D_{\boldsymbol{F}'}$. Furthermore, parameter $\epsilon'$ is a continuous non-decreasing function of $\epsilon$.\newline
2) When $\epsilon$ changes from 0 to 1, point $Z_{\boldsymbol{F}''}$ moves on the line segment $\overline{Z_{\boldsymbol{F}}Z_{\boldsymbol{F}'}}$ from $Z_{\boldsymbol{F}}$ to $Z_{\boldsymbol{F}'}$.
\label{lemma_linearcombination}
\end{lemma}
\begin{IEEEproof}
In the following, the two conclusions of the lemma will be proven one by one.

1) According to the definition of $\boldsymbol{H}_{\boldsymbol{F}}$ and $\boldsymbol{p}_{\boldsymbol{F}}$, we have that if $\boldsymbol{F}''=(1-\epsilon)\boldsymbol{F}+\epsilon\boldsymbol{F}'$, then $\boldsymbol{H}_{\boldsymbol{F}''}=(1-\epsilon)\boldsymbol{H}_{\boldsymbol{F}}+\epsilon\boldsymbol{H}_{\boldsymbol{F}'}$ and $\boldsymbol{p}_{\boldsymbol{F}''}=(1-\epsilon)\boldsymbol{p}_{\boldsymbol{F}}+\epsilon\boldsymbol{p}_{\boldsymbol{F}'}$. Set $\Delta\boldsymbol{H}=\boldsymbol{H}_{\boldsymbol{F}'}-\boldsymbol{H}_{\boldsymbol{F}}$ and $\Delta\boldsymbol{p}=\boldsymbol{p}_{\boldsymbol{F}'}-\boldsymbol{p}_{\boldsymbol{F}}$. Since $\boldsymbol{F}$ and $\boldsymbol{F}'$ are different only in the $(q+1)$th row, it can be derived that the $(q+1)$th column of $\Delta \boldsymbol{H}$ is the only column that can contain non-zero elements, and the $(q+1)$th element of $\Delta \boldsymbol{p}$ is its only non-zero element. Therefore $\Delta \boldsymbol{H}$ can be expressed as $\left[\boldsymbol{0},\cdots,\boldsymbol{\delta}_q,\cdots,\boldsymbol{0}\right]$, where $\boldsymbol{\delta}_q$ is its $(q+1)$th column, and $\Delta \boldsymbol{p}$ can be expressed as $\left[0,\cdots,\zeta_q,\cdots,0\right]^T$, where $\zeta_q$ is its $(q+1)$th element. Based on this, we set
\begin{align}
\boldsymbol{H}_{\boldsymbol{F}}^{-1}=\left[
\boldsymbol{h}_0^T, \boldsymbol{h}_1^T, \cdots, \boldsymbol{h}_Q^T
\right]^T.
\end{align}
Hence
\begin{align}
(\boldsymbol{H}_{\boldsymbol{F}}^{-1}\Delta\boldsymbol{H})\boldsymbol{H}_{\boldsymbol{F}}^{-1}=\left[
\begin{array}{c}
(\boldsymbol{h}_0^T\boldsymbol{\delta}_q)\boldsymbol{h}_q^T\\
(\boldsymbol{h}_1^T\boldsymbol{\delta}_q)\boldsymbol{h}_q^T\\
\vdots\\
(\boldsymbol{h}_Q^T\boldsymbol{\delta}_q)\boldsymbol{h}_q^T
\end{array}
\right].
\end{align}

By mathematical induction, we can prove that for $i\ge 1$,
\begin{align}
&(\boldsymbol{H}_{\boldsymbol{F}}^{-1}\Delta\boldsymbol{H})^{i}\boldsymbol{H}_{\boldsymbol{F}}^{-1}\nonumber\\
=&\left[
\begin{array}{c}
(\boldsymbol{h}_0^T\boldsymbol{\delta}_q)(\boldsymbol{h}_q^T\boldsymbol{\delta}_q)^{i-1}\boldsymbol{h}_q^T\\
(\boldsymbol{h}_1^T\boldsymbol{\delta}_q)(\boldsymbol{h}_q^T\boldsymbol{\delta}_q)^{i-1}\boldsymbol{h}_q^T\\
\vdots\\
(\boldsymbol{h}_Q^T\boldsymbol{\delta}_q)(\boldsymbol{h}_q^T\boldsymbol{\delta}_q)^{i-1}\boldsymbol{h}_q^T
\end{array}
\right]\\
=&(\boldsymbol{h}_q^T\boldsymbol{\delta}_q)^{i-1}(\boldsymbol{H}_{\boldsymbol{F}}^{-1}\Delta\boldsymbol{H})\boldsymbol{H}_{\boldsymbol{F}}^{-1}
\end{align}
and
\begin{align}
&\Delta\boldsymbol{p}^T\boldsymbol{H}_{\boldsymbol{F}}^{-1}(\boldsymbol{H}_{\boldsymbol{F}}^{-1}\Delta\boldsymbol{H})^{i-1}\nonumber\\
=&\zeta_q(\boldsymbol{h}_q^T\boldsymbol{\delta}_q)^{i-1}\boldsymbol{h}_q^T.
\end{align}

Therefore,
\begin{align}
&(\boldsymbol{H}_{\boldsymbol{F}}+\epsilon\Delta\boldsymbol{H})^{-1}\nonumber\\
=&\sum_{i=0}^{+\infty} (-\epsilon)^{i}(\boldsymbol{H}_{\boldsymbol{F}}^{-1}\Delta\boldsymbol{H})^{i}\boldsymbol{H}_{\boldsymbol{F}}^{-1}\\
=&\boldsymbol{H}_{\boldsymbol{F}}^{-1}
+\sum_{i=1}^{+\infty}(-\epsilon)^i(\boldsymbol{h}_q^T\boldsymbol{\delta}_q)^{i-1}(\boldsymbol{H}_{\boldsymbol{F}}^{-1}\Delta\boldsymbol{H})\boldsymbol{H}_{\boldsymbol{F}}^{-1}.
\end{align}

We have $P_{\boldsymbol{F}}=\boldsymbol{p}_{\boldsymbol{F}}^T\boldsymbol{H}_{\boldsymbol{F}}^{-1}\boldsymbol{c}$ and $D_{\boldsymbol{F}}=\frac{1}{E_a}\boldsymbol{d}^T\boldsymbol{H}_{\boldsymbol{F}}^{-1}\boldsymbol{c}$. Hence
\begin{align}
&\frac{P_{\boldsymbol{F}''}-P_{\boldsymbol{F}}}{P_{\boldsymbol{F}'}-P_{\boldsymbol{F}}}\nonumber\\
=&\frac{(\boldsymbol{p}_{\boldsymbol{F}}+\epsilon\Delta\boldsymbol{p})^T(\boldsymbol{H}_{\boldsymbol{F}}+\epsilon\Delta\boldsymbol{H})^{-1}\boldsymbol{c}-\boldsymbol{p}_{\boldsymbol{F}}^T\boldsymbol{H}_{\boldsymbol{F}}^{-1}\boldsymbol{c}}{(\boldsymbol{p}_{\boldsymbol{F}}+\Delta\boldsymbol{p})^T(\boldsymbol{H}_{\boldsymbol{F}}+\Delta\boldsymbol{H})^{-1}\boldsymbol{c}-\boldsymbol{p}_{\boldsymbol{F}}^T\boldsymbol{H}_{\boldsymbol{F}}^{-1}\boldsymbol{c}}\\
=&\frac{
\begin{array}{c}
\boldsymbol{p}_{\boldsymbol{F}}^T\left[(\boldsymbol{H}_{\boldsymbol{F}}+\epsilon\Delta\boldsymbol{H})^{-1}-\boldsymbol{H}_{\boldsymbol{F}}^{-1}\right]\boldsymbol{c}\\
+\epsilon\Delta\boldsymbol{p}^T(\boldsymbol{H}_{\boldsymbol{F}}+\epsilon\Delta\boldsymbol{H})^{-1}\boldsymbol{c}
\end{array}
}
{
\begin{array}{c}
\boldsymbol{p}_{\boldsymbol{F}}^T\left[(\boldsymbol{H}_{\boldsymbol{F}}+\Delta\boldsymbol{H})^{-1}-\boldsymbol{H}_{\boldsymbol{F}}^{-1}\right]\boldsymbol{c}\\
+\Delta\boldsymbol{p}^T(\boldsymbol{H}_{\boldsymbol{F}}+\Delta\boldsymbol{H})^{-1}\boldsymbol{c}
\end{array}
}\\
=&\frac{
\begin{array}{c}
\boldsymbol{p}_{\boldsymbol{F}}^T\left[\sum_{i=1}^{+\infty}(-\epsilon)^i(\boldsymbol{h}_q^T\boldsymbol{\delta}_q)^{i-1}(\boldsymbol{H}_{\boldsymbol{F}}^{-1}\Delta\boldsymbol{H})\boldsymbol{H}_{\boldsymbol{F}}^{-1}\right]\boldsymbol{c}\\
-\Delta\boldsymbol{p}^T\left[\sum_{i=1}^{+\infty} (-\epsilon)^{i}(\boldsymbol{H}_{\boldsymbol{F}}^{-1}\Delta\boldsymbol{H})^{i-1}\boldsymbol{H}_{\boldsymbol{F}}^{-1}\right]\boldsymbol{c}
\end{array}
}
{
\begin{array}{c}
\boldsymbol{p}_{\boldsymbol{F}}^T\left[\sum_{i=1}^{+\infty}(-1)^i(\boldsymbol{h}_q^T\boldsymbol{\delta}_q)^{i-1}(\boldsymbol{H}_{\boldsymbol{F}}^{-1}\Delta\boldsymbol{H})\boldsymbol{H}_{\boldsymbol{F}}^{-1}\right]\boldsymbol{c}\\
-\Delta\boldsymbol{p}^T\left[\sum_{i=1}^{+\infty} (-1)^{i}(\boldsymbol{H}_{\boldsymbol{F}}^{-1}\Delta\boldsymbol{H})^{i-1}\boldsymbol{H}_{\boldsymbol{F}}^{-1}\right]\boldsymbol{c}
\end{array}
}\\
=&\frac{
\begin{array}{c}
\sum_{i=1}^{+\infty}(-\epsilon)^i(\boldsymbol{h}_q^T\boldsymbol{\delta}_q)^{i-1}\boldsymbol{p}_{\boldsymbol{F}}^T(\boldsymbol{H}_{\boldsymbol{F}}^{-1}\Delta\boldsymbol{H})\boldsymbol{H}_{\boldsymbol{F}}^{-1}\boldsymbol{c}\\
-\sum_{i=1}^{+\infty} (-\epsilon)^{i}\zeta_q(\boldsymbol{h}_q^T\boldsymbol{\delta}_q)^{i-1}\boldsymbol{h}_q^T\boldsymbol{c}
\end{array}
}
{
\begin{array}{c}
\sum_{i=1}^{+\infty}(-1)^i(\boldsymbol{h}_q^T\boldsymbol{\delta}_q)^{i-1}\boldsymbol{p}_{\boldsymbol{F}}^T(\boldsymbol{H}_{\boldsymbol{F}}^{-1}\Delta\boldsymbol{H})\boldsymbol{H}_{\boldsymbol{F}}^{-1}\boldsymbol{c}\\
-\sum_{i=1}^{+\infty} (-1)^{i}\zeta_q(\boldsymbol{h}_q^T\boldsymbol{\delta}_q)^{i-1}\boldsymbol{h}_q^T\boldsymbol{c}
\end{array}
}\\
=&\frac{\sum_{i=1}^{+\infty}(-\epsilon)^i(\boldsymbol{h}_q^T\boldsymbol{\delta}_q)^{i-1}}
{\sum_{i=1}^{+\infty}(-1)^i(\boldsymbol{h}_q^T\boldsymbol{\delta}_q)^{i-1}}\\
=&\frac{\epsilon+\epsilon\boldsymbol{h}_q^T\boldsymbol{\delta}_q}{1+\epsilon\boldsymbol{h}_q^T\boldsymbol{\delta}_q}
\end{align}
and
\begin{align}
&\frac{D_{\boldsymbol{F}''}-D_{\boldsymbol{F}}}{D_{\boldsymbol{F}'}-D_{\boldsymbol{F}}}\nonumber\\
=&\frac{\boldsymbol{d}^T(\boldsymbol{H}_{\boldsymbol{F}}+\epsilon\Delta\boldsymbol{H})^{-1}\boldsymbol{c}-\boldsymbol{d}^T\boldsymbol{H}_{\boldsymbol{F}}^{-1}\boldsymbol{c}}{\boldsymbol{d}^T(\boldsymbol{H}_{\boldsymbol{F}}+\Delta\boldsymbol{H})^{-1}\boldsymbol{c}-\boldsymbol{d}^T\boldsymbol{H}_{\boldsymbol{F}}^{-1}\boldsymbol{c}}\\
=&\frac{\boldsymbol{d}^T(\sum_{i=1}^{+\infty}(-\epsilon)^i(\boldsymbol{h}_q^T\boldsymbol{\delta}_q)^{i-1}(\boldsymbol{H}_{\boldsymbol{F}}^{-1}\Delta\boldsymbol{H})\boldsymbol{H}_{\boldsymbol{F}}^{-1})\boldsymbol{c}}{\boldsymbol{d}^T(\sum_{i=1}^{+\infty}(-1)^i(\boldsymbol{h}_q^T\boldsymbol{\delta}_q)^{i-1}(\boldsymbol{H}_{\boldsymbol{F}}^{-1}\Delta\boldsymbol{H})\boldsymbol{H}_{\boldsymbol{F}}^{-1})\boldsymbol{c}}\\
=&\frac{\sum_{i=1}^{+\infty}(-\epsilon)^i(\boldsymbol{h}_q^T\boldsymbol{\delta}_q)^{i-1}}{\sum_{i=1}^{+\infty}(-1)^i(\boldsymbol{h}_q^T\boldsymbol{\delta}_q)^{i-1}}\\
=&\frac{\epsilon+\epsilon\boldsymbol{h}_q^T\boldsymbol{\delta}_q}{1+\epsilon\boldsymbol{h}_q^T\boldsymbol{\delta}_q}.
\end{align}
Hence $\frac{P_{\boldsymbol{F}''}-P_{\boldsymbol{F}}}{P_{\boldsymbol{F}'}-P_{\boldsymbol{F}}}=\frac{D_{\boldsymbol{F}''}-D_{\boldsymbol{F}}}{D_{\boldsymbol{F}'}-D_{\boldsymbol{F}}}=\frac{\epsilon+\epsilon\boldsymbol{h}_q^T\boldsymbol{\delta}_q}{1+\epsilon\boldsymbol{h}_q^T\boldsymbol{\delta}_q}=\epsilon'$, so that $P_{\boldsymbol{F}''}=(1-\epsilon')P_{\boldsymbol{F}}+\epsilon' P_{\boldsymbol{F}'}$ and $D_{\boldsymbol{F}''}=(1-\epsilon')D_{\boldsymbol{F}}+\epsilon' D_{\boldsymbol{F}'}$. Furthermore, it can be seen that $\epsilon'=\frac{\epsilon+\epsilon\boldsymbol{h}_q^T\boldsymbol{\delta}_q}{1+\epsilon\boldsymbol{h}_q^T\boldsymbol{\delta}_q}$ is a continuous nondecreasing function.

2) From the first part, we proved $\frac{P_{\boldsymbol{F}''}-P_{\boldsymbol{F}}}{P_{\boldsymbol{F}'}-P_{\boldsymbol{F}}}=\frac{D_{\boldsymbol{F}''}-D_{\boldsymbol{F}}}{D_{\boldsymbol{F}'}-D_{\boldsymbol{F}}}=\epsilon'$ and $\epsilon'$ is a continuous non-decreasing function of $\epsilon$. When $\epsilon=0$, we have $\epsilon'=0$. When $\epsilon=1$, we have $\epsilon'=1$. Therefore when $\epsilon$ changes from 0 to 1, the point $(P_{\boldsymbol{F}''},D_{\boldsymbol{F}''})$ moves on the line segment from $(P_{\boldsymbol{F}},D_{\boldsymbol{F}})$ to $(P_{\boldsymbol{F}'},D_{\boldsymbol{F}'})$. The slope of the line can be expressed as
\begin{align}
&\frac{D_{\boldsymbol{F}'}-D_{\boldsymbol{F}}}{P_{\boldsymbol{F}'}-P_{\boldsymbol{F}}}=\frac{
\frac{1}{E_a}\boldsymbol{d}^T(\boldsymbol{H}_{\boldsymbol{F}}+\Delta\boldsymbol{H})^{-1}\boldsymbol{c}-\frac{1}{E_a}\boldsymbol{d}^T\boldsymbol{H}_{\boldsymbol{F}}^{-1}\boldsymbol{c}
}{
(\boldsymbol{p}_{\boldsymbol{F}}+\Delta\boldsymbol{p})^T(\boldsymbol{H}_{\boldsymbol{F}}+\Delta\boldsymbol{H})^{-1}\boldsymbol{c}-\boldsymbol{p}_{\boldsymbol{F}}^T\boldsymbol{H}_{\boldsymbol{F}}^{-1}\boldsymbol{c}
}\\
=&\frac{\frac{1}{E_a}\boldsymbol{d}^T\boldsymbol{H}_{\boldsymbol{F}}^{-1}\Delta\boldsymbol{H}\boldsymbol{H}_{\boldsymbol{F}}^{-1}\boldsymbol{c}}
{\boldsymbol{p}_{\boldsymbol{F}}^T\boldsymbol{H}_{\boldsymbol{F}}^{-1}\Delta\boldsymbol{H}\boldsymbol{H}_{\boldsymbol{F}}^{-1}\boldsymbol{c}-\zeta_q\boldsymbol{h}_q^T\boldsymbol{c}}=\frac{\boldsymbol{d}^T\boldsymbol{H}_{\boldsymbol{F}}^{-1}\boldsymbol{\delta}_q}
{E_a (\boldsymbol{p}_{\boldsymbol{F}}^T\boldsymbol{H}_{\boldsymbol{F}}^{-1}\boldsymbol{\delta}_q-\zeta_q)}.\label{slope}
\end{align}
\end{IEEEproof}

\subsection{Properties of set $\mathcal{R}$}
In this subsection, we will prove that $\mathcal{R}$, the set of all feasible points in the delay-power plane, is a convex polygon whose vertices are all obtained by deterministic scheduling policies. Moreover, the policies corresponding to adjacent vertices of $\mathcal{R}$ take different actions in only one state.

\begin{figure*}[t]
\centering
\subfloat[A Convex Basic Polygon in the Normal Shape]{\includegraphics[width=0.9\columnwidth]{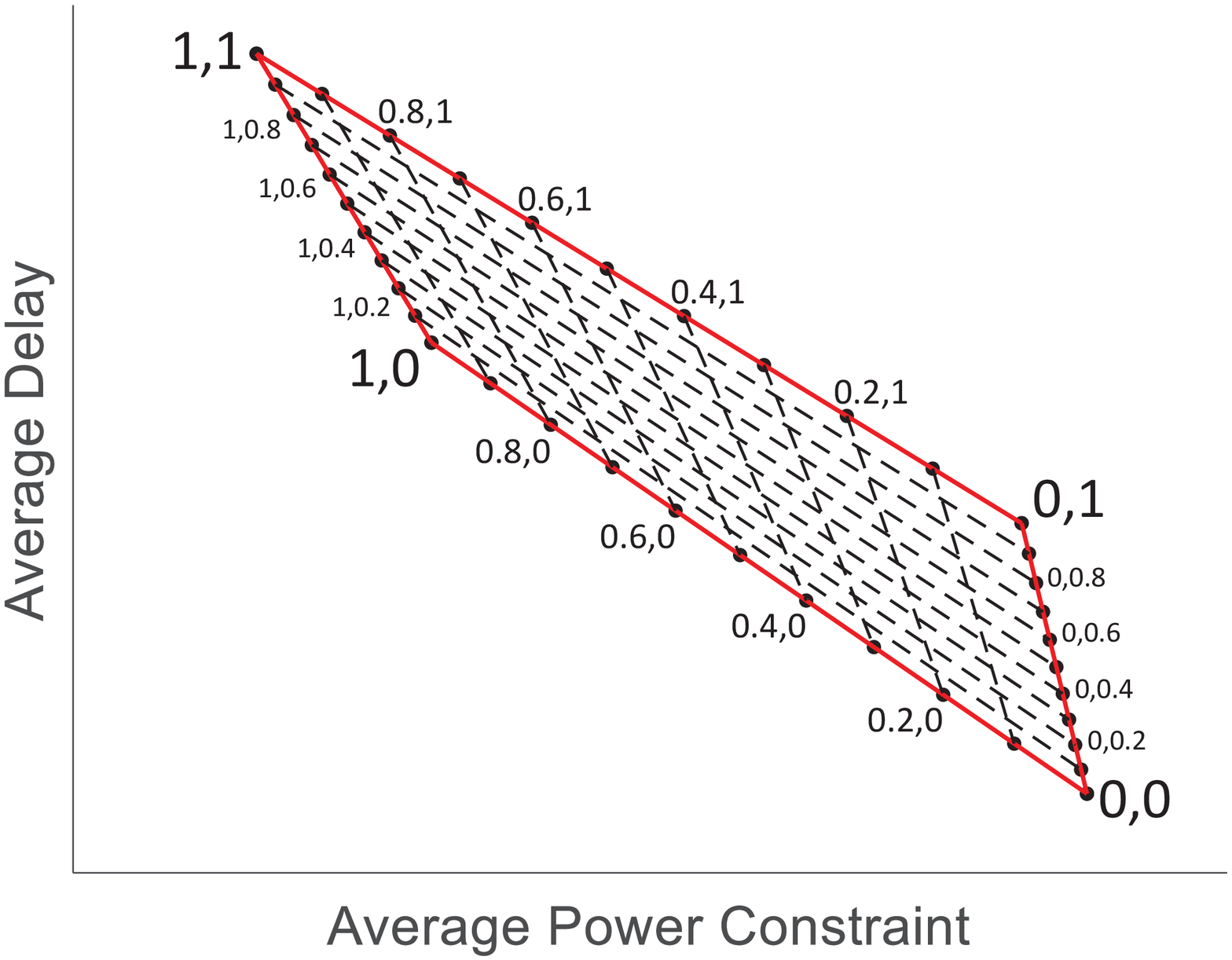}
\label{fig_basic_1}}
\subfloat[A Nonconvex Basic Polygon in the Boomerang Shape]{\includegraphics[width=0.9\columnwidth]{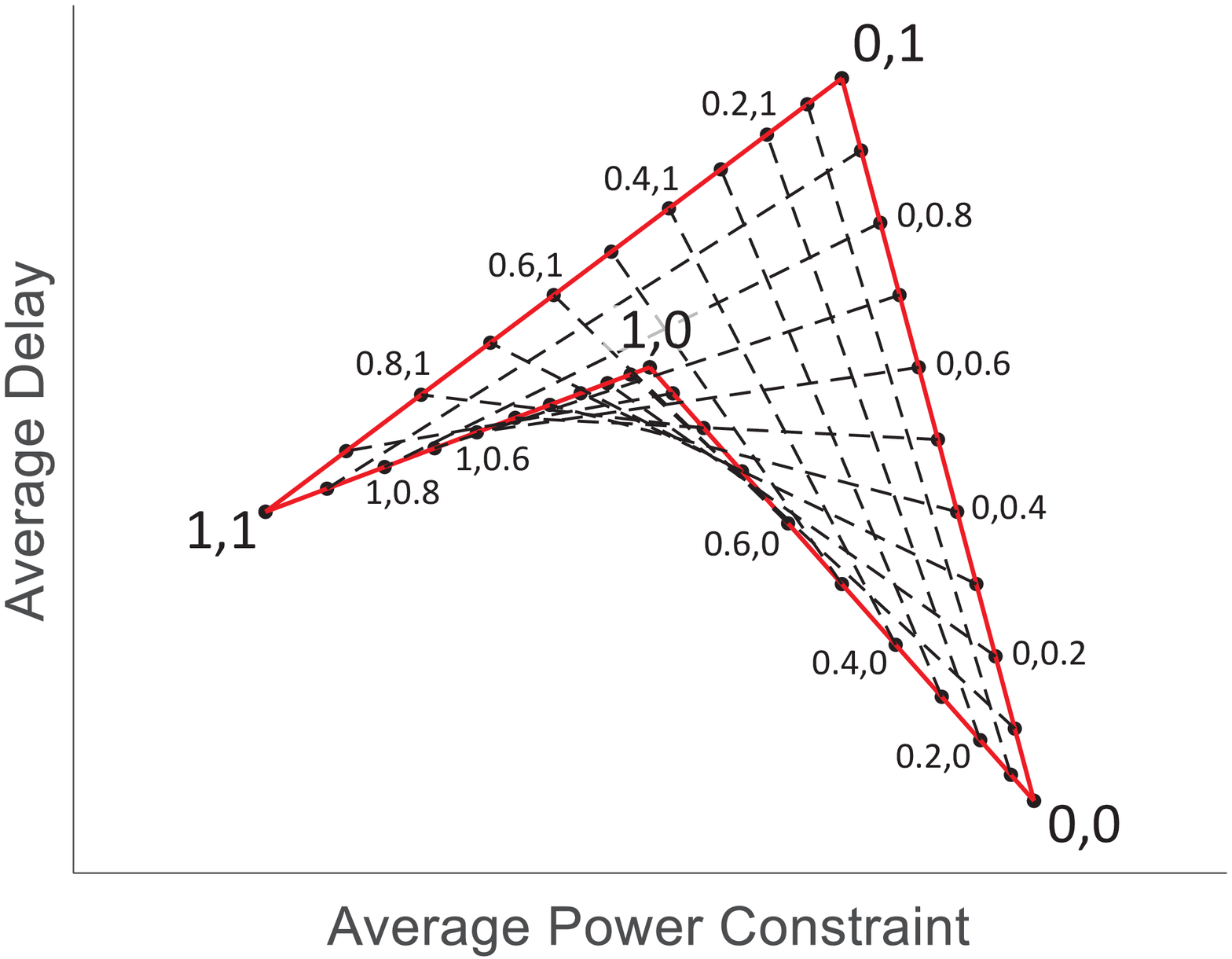}
\label{fig_basic_2}}\\
\subfloat[A NonConvex Basic Polygon in the Butterfly Shape]{\includegraphics[width=0.9\columnwidth]{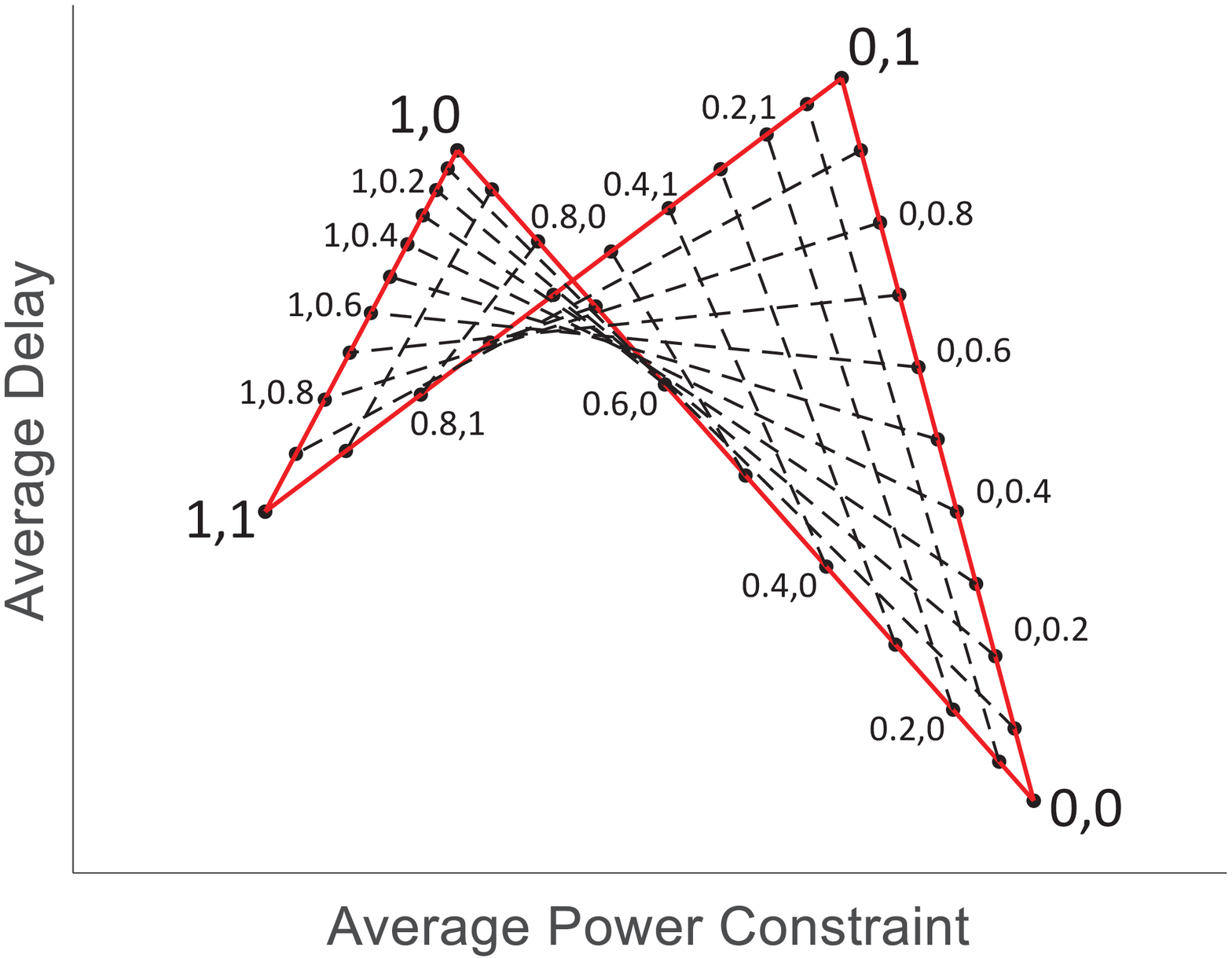}
\label{fig_basic_3}}
\subfloat[A Nonconvex Basic Polygon in the Slender Butterfly Shape]{\includegraphics[width=0.9\columnwidth]{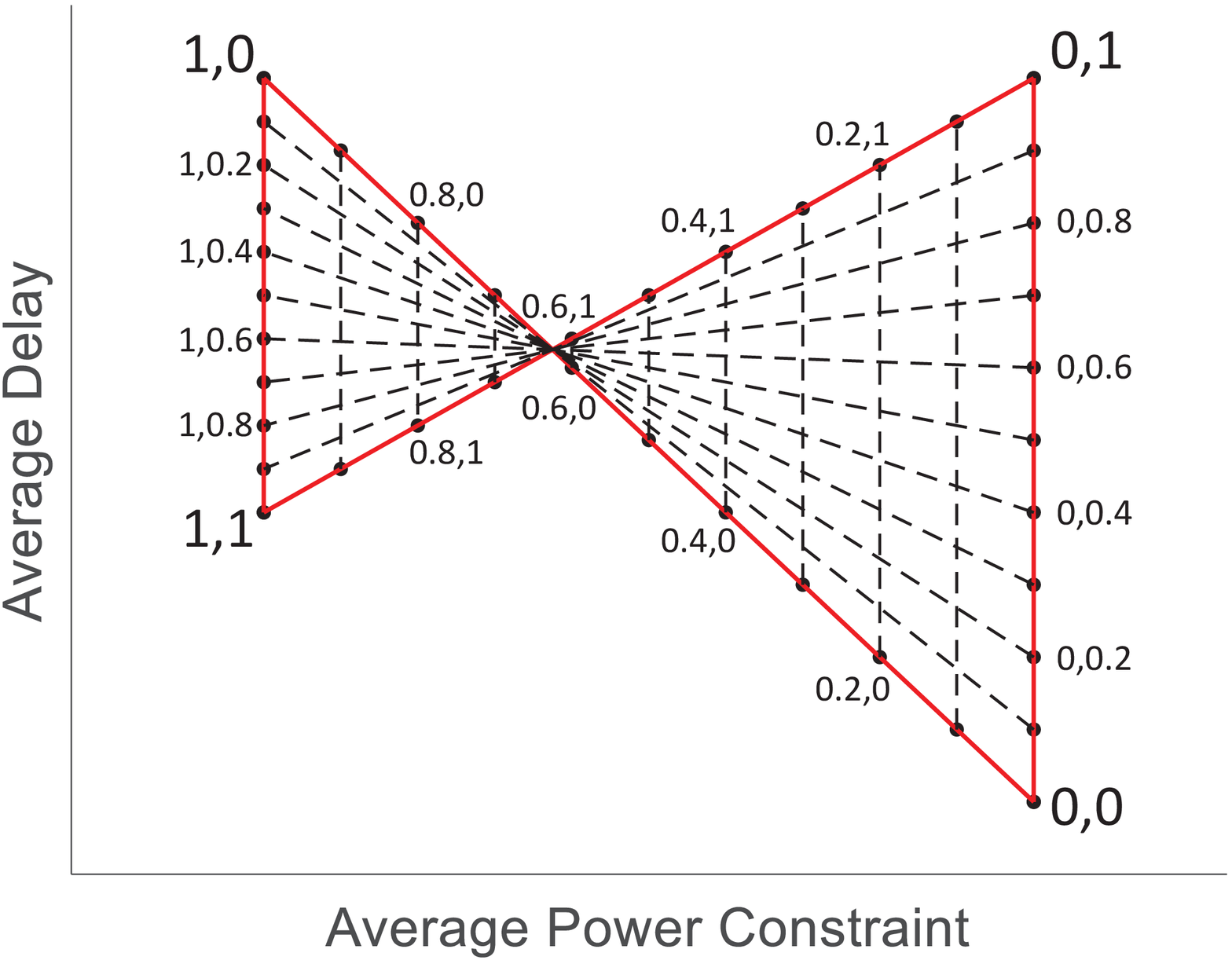}
\label{fig_basic_4}}
\caption{Demonstration for Basic Polygons}
\label{fig_basic}
\end{figure*}

Define $\mathcal{C}=\textbf{conv } \{Z_{\boldsymbol{F}} | \boldsymbol{F} \in \mathcal{F}_D \}$ as the convex hull of points corresponding to deterministic scheduling policies in the delay-power plane. Hence we will show that $\mathcal{R}$ is a convex polygon whose vertices are all obtained by deterministic scheduling policies by proving $\mathcal{R}=\mathcal{C}$.

The proof is made up of three parts. In Part I, we will prove $\mathcal{R}\subseteq\mathcal{C}$ by the construction method. Part II is the most difficult part. We will first define the concepts of basic polygons and compound polygons, then prove their convexity, based on which $\mathcal{R}\supseteq\mathcal{C}$ can be proven. By combining the results from Part I and II, we will have $\mathcal{R}=\mathcal{C}$. Finally, in Part III, it will be shown that policies corresponding to adjacent vertices of $\mathcal{R}$ are different in only one state.

\textbf{Part I. Prove $\mathcal{R}\subseteq\mathcal{C}$}

For any probabilistic policy $\boldsymbol{F}$ where $0<f_{q^*,s^*}<1$, we construct
\begin{align}
\boldsymbol{F}'=
\begin{cases}
f'_{q,s}=1 & q=q^*,s=s^*\\
f'_{q,s}=0 & q=q^*,s \neq s^*\\
f'_{q,s}=f_{q,s} & \text{else}
\end{cases}
\end{align}
and
\begin{align}
\boldsymbol{F}''=
\begin{cases}
f''_{q,s}=0 & q=q^*,s=s^*\\
f''_{q,s}=\frac{f_{q,s}}{1-f_{q^*,s^*}} & q=q^*,s \neq s^*\\
f''_{q,s}=f_{q,s} & \text{else}.
\end{cases}
\end{align}

Since $0 \le \frac{f_{q,s}}{1-f_{q^*,s^*}} \le 1$, and the fact that whenever $f_{q,s}=0$, it must holds that $f'_{q,s}=f''_{q,s}=0$, we can conclude that policies $\boldsymbol{F}'$ and $\boldsymbol{F}''$ are feasible. It can be seen that $\boldsymbol{F}=f_{q^*,s^*}\boldsymbol{F}'+(1-f_{q^*,s^*})\boldsymbol{F}''$. Since $\boldsymbol{F}$ is a convex combination of policy $\boldsymbol{F}'$ and policy $\boldsymbol{F}''$, also $\boldsymbol{F}'$ and $\boldsymbol{F}''$ are different only in the $(q^*+1)$th row, from Lemma \ref{lemma_linearcombination}, we know that $Z_{\boldsymbol{F}}$ is a convex combination of $Z_{\boldsymbol{F}'}$ and $Z_{\boldsymbol{F}''}$. Note that $f'_{q^*,s^*}$ and $f''_{q^*,s^*}$ are integers. Also, in matrices $\boldsymbol{F}'$ and $\boldsymbol{F}''$, no new decimal elements are going to be introduced. Hence in finite steps, the point $Z_{\boldsymbol{F}}$ can be expressed as the convex combination of points corresponding to deterministic scheduling policies. That is to say $Z_{\boldsymbol{F}}\in\mathcal{C}$. From the arbitrariness of $\boldsymbol{F}$, we have $\mathcal{R}\subseteq\mathcal{C}$.

\textbf{Part II. Prove $\mathcal{R}\supseteq\mathcal{C}$}

In this part, we will begin with the concepts of basic polygons and compound polygons in Part II.0. Then we will prove that basic polygons and compound polygons are convex in Part II.1 and Part II.2 respectively. Based on the above results, we will prove $\mathcal{R}\supseteq\mathcal{C}$ in Part II.3.

\textbf{Part II.0 The Concepts of Basic Polygons and Compound Polygons}

\begin{figure}[t]
\centering
\subfloat[A Convex Compound Polygon]{\includegraphics[width=0.6\columnwidth]{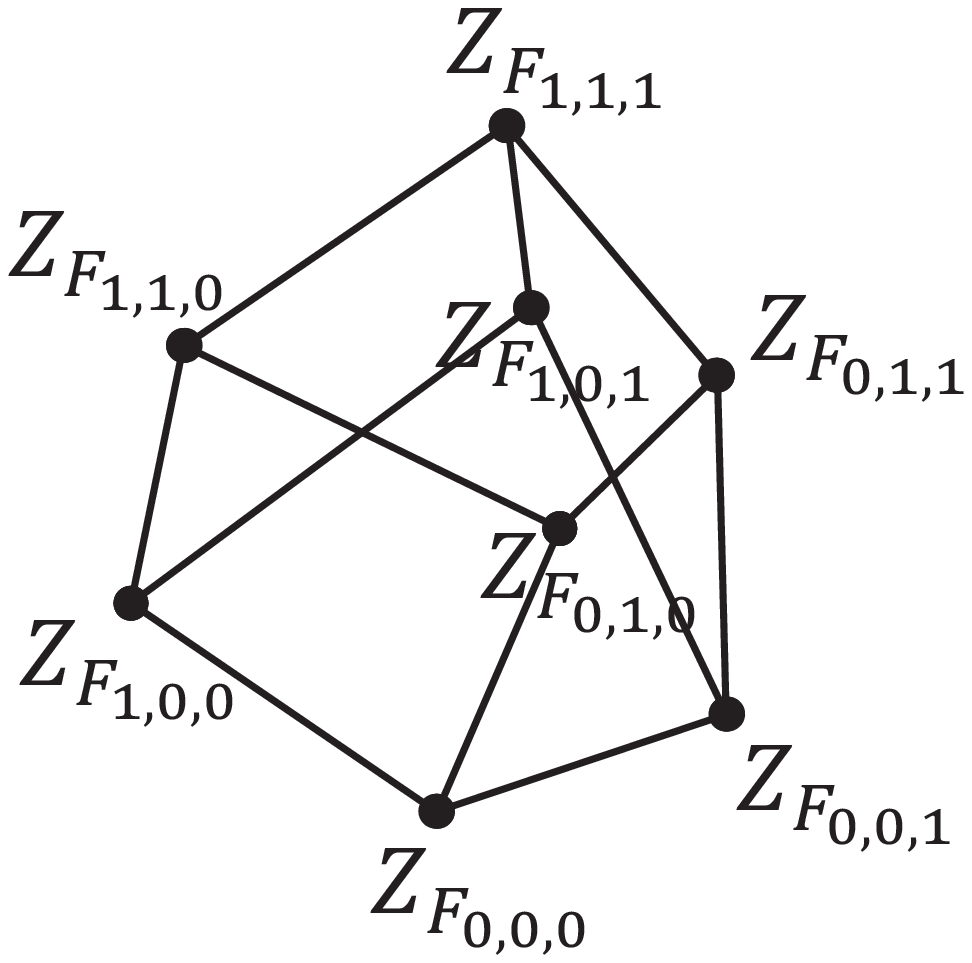}
\label{fig_convex_1}}
\subfloat[A Nonconvex Compound Polygon]{\includegraphics[width=0.4\columnwidth]{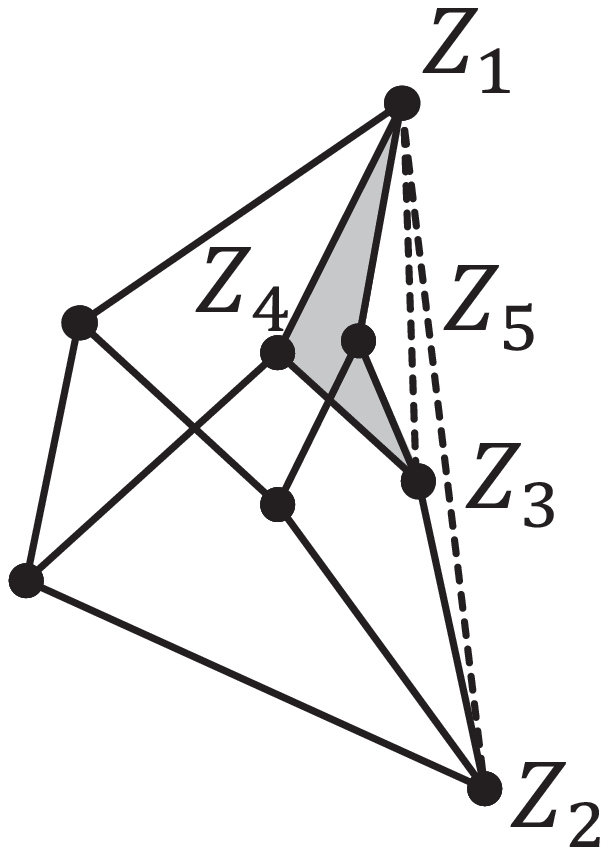}
\label{fig_convex_2}}
\caption{Demonstration for Compound Polygons}
\label{fig_convex}
\end{figure}

For two deterministic policies $\boldsymbol{F}$ and $\boldsymbol{F}'$ which are different in $K$ states, namely $q_1,\cdots,q_K$, define
$\boldsymbol{F}_{b_1,b_2,\cdots,b_K}(q,:)=\begin{cases}
(1-b_k)\boldsymbol{F}(q,:)+b_k\boldsymbol{F}'(q,:) & q=q_k,\\
\boldsymbol{F}(q,:) & q\neq q_1,\cdots,q_K,
\end{cases}$
where $0 \le b_k \le 1$ for all $k$. Thus $\boldsymbol{F}_{0,0,\cdots,0}=\boldsymbol{F}$, and $\boldsymbol{F}_{1,1,\cdots,1}=\boldsymbol{F}'$. With more $b_k$ close to 0, the policy is more like $\boldsymbol{F}$. With more $b_k$ close to 1, the policy is more like $\boldsymbol{F}'$. For policies $\boldsymbol{F}_{b_1,\cdots,b_k,\cdots,b_K}$ and $\boldsymbol{F}_{b_1,\cdots,b_k',\cdots,b_K}$ where $b_k \neq b_k'$, since they are different in only one state, according to Lemma \ref{lemma_linearcombination}, the delay-power point corresponding to their convex combination $Z_{\epsilon \boldsymbol{F}_{b_1,\cdots,b_k,\cdots,b_K} + (1-\epsilon)\boldsymbol{F}_{b_1,\cdots,b_k',\cdots,b_K}}$ is the convex combination of $Z_{\boldsymbol{F}_{b_1,\cdots,b_k,\cdots,b_K}}$ and $Z_{\boldsymbol{F}_{b_1,\cdots,b_k',\cdots,b_K}}$. However, for two policies which are different in more than one state, the delay-power points corresponding to policies of their convex combination are not necessarily the convex combination of delay-power points corresponding to themselves. Therefore, we introduce the concept of generated polygon for the delay-power region of point set corresponding to policies which are convex combinations of the two policies. We plot $Z_{\boldsymbol{F}_{b_1,\cdots,b_K}}$, where $b_k=0$ or $1$ for all $1\le k \le K$, and connect the points whose corresponding policies are different in only one state. Therefore any point on any line segment can be obtained by a certain policy. We define the figure as a polygon generated by $\boldsymbol{F}$ and $\boldsymbol{F}'$. The red polygon in \figurename~\ref{fig_basic_1} and the polygon in \figurename~\ref{fig_convex_1} are demonstrations where $\boldsymbol{F}$ and $\boldsymbol{F}'$ are different in 2 and 3 states respectfully. If $K=2$, we call the polygon a basic polygon. If $K>2$, we call it a compound polygon. As demonstrated in \figurename~\ref{fig_convex_1}, a compound polygon contains multiple basic polygons.

\textbf{Part II.1 Prove a Basic Polygon is Convex and Any Point Inside a Basic Polygon can be Obtained by a Policy}

For better visuality, in \figurename~\ref{fig_basic}, we simplify the notation $Z_{\boldsymbol{F}_{b_1,b_2}}$ as $b_1,b_2$. By considering all possible relative positions of $Z_{\boldsymbol{F}_{0,0}}$, $Z_{\boldsymbol{F}_{0,1}}$, $Z_{\boldsymbol{F}_{1,0}}$, and $Z_{\boldsymbol{F}_{1,1}}$, there are 3 possible shapes of basic polygons in total, as shown in \figurename~\ref{fig_basic_1}-\ref{fig_basic_3} respectfully. We name them as the normal shape, the boomerang shape, and the butterfly shape. The degenerate cases such as triangles, line segments and points are considered included in the above three cases. Besides $\boldsymbol{F}_{b_1,b_2}$ with integral $b_1,b_2$ and the line segments connecting them, in the figures we also plot the points corresponding to policy $\boldsymbol{F}_{b_1,b_2}$ where one of $b_1,b_2$ is integer and the other one is decimal. We connect the points corresponding to policies which have the same $b_1$ or $b_2$ with dashed lines. As demonstrated in \figurename~\ref{fig_basic}, we draw line segments $\overline{Z_{\boldsymbol{F}_{b_1,0}}Z_{\boldsymbol{F}_{b_1,1}}}$ where $b_1=0.1,0.2,\cdots,0.9$ and $\overline{Z_{\boldsymbol{F}_{0,b_2}}Z_{\boldsymbol{F}_{1,b_2}}}$ where $b_2=0.1,0.2,\cdots,0.9$. For any specific $b_1$ and $b_2$, the point $Z_{\boldsymbol{F}_{b_1,b_2}}$ should be on both $\overline{Z_{\boldsymbol{F}_{b_1,0}}Z_{\boldsymbol{F}_{b_1,1}}}$ and $\overline{Z_{\boldsymbol{F}_{0,b_2}}Z_{\boldsymbol{F}_{1,b_2}}}$. Because of the existence of $Z_{\boldsymbol{F}_{b_1,b_2}}$, line segments $\overline{Z_{\boldsymbol{F}_{b_1,0}}Z_{\boldsymbol{F}_{b_1,1}}}$ and $\overline{Z_{\boldsymbol{F}_{0,b_2}}Z_{\boldsymbol{F}_{1,b_2}}}$ should always have an intersection point for any specific $b_1$ and $b_2$. However, if there exist line segments outside the polygon, there exist $b_1$ and $b_2$ whose line segments don't intersect. Therefore, in the boomerang shape, there will always exist $b_1$ and $b_2$ whose line segments don't intersect. In the butterfly shape, there will exist $b_1$ and $b_2$ whose line segments don't intersect except the case that all the line segments are inside the basic polygon, as shown in \figurename~\ref{fig_basic_4}, which is named as the slender butterfly shape. In the slender butterfly shape, there exists a specific $b_1^*$ such that $\overline{Z_{\boldsymbol{F}_{b_1^*,0}}Z_{\boldsymbol{F}_{b_1^*,1}}}$ degenerates into a point, or there exists a specific $b_2^*$ such that $\overline{Z_{\boldsymbol{F}_{0,b_2^*}}Z_{\boldsymbol{F}_{1,b_2^*}}}$ degenerates into a point. Without loss of generality, we assume it is the $b_1^*$ case. It means that under policy $\boldsymbol{F}_{b_1^*,b_2}$, state $q_2$, the state corresponding to $b_2$, is a transient state. For $b_1\in(b_1^*-\epsilon,b_1^*+\epsilon)$ when $\epsilon$ is small enough, the Markov chain applying policy $F_{b_1,b_2}$ also has $q_2$ as a transient state, therefore $\overline{Z_{\boldsymbol{F}_{b_1,0}}Z_{\boldsymbol{F}_{b_1,1}}}$ also degenerates into a point. Thus $\overline{Z_{\boldsymbol{F}_{0,0}}Z_{\boldsymbol{F}_{1,0}}}$ and $\overline{Z_{\boldsymbol{F}_{0,1}}Z_{\boldsymbol{F}_{1,1}}}$ overlap, which means the slender butterfly shape always degenerates to a line segment, which can also be considered as a normal shape. Since the normal shape is the only possible shape of a basic polygon, the basic polygon is convex. Since the transition from the point $Z_{\boldsymbol{F}_{0,0}}$ to $Z_{\boldsymbol{F}_{1,1}}$ is termwise monotone and continuous, every point inside the basic polygon can be obtained by a policy.

\textbf{Part II.2 Prove a Compound Polygon is Convex}

For any two deterministic policies $\boldsymbol{F}$ and $\boldsymbol{F}'$, if the compound polygon generated by them is not convex, then there must exist two vertices whose connecting line is outside the compound polygon, as demonstrated by $\overline{Z_1 Z_2}$ in \figurename~\ref{fig_convex_2}. Therefore, there must also exist two vertices who are connecting to the same point and their connecting line is outside the compound polygon, as demonstrated by $\overline{Z_1 Z_3}$. The policies corresponding to these two vertices must be different in only two states based on previous conclusions, therefore there must be a basic polygon generated by them, as demonstrated in \figurename~\ref{fig_convex_2} by the filled polygon. Since $\overline{Z_1 Z_3}$ is outside the compound polygon, it is for sure that $\overline{Z_1 Z_3}$ is outside the basic polygon too. This is impossible because basic polygons are always convex. Hence we can conclude that all generated compound polygons are convex.

\textbf{Part II.3 Prove $\mathcal{R}\supseteq\mathcal{C}$}

For arbitrary point $C\in\mathcal{C}$, it will fall into one of the compound polygons, because otherwise, there will be at least one point corresponding to a deterministic policy outside any compound polygons. All compound polygons can be covered by basic polygons, therefore $C$ is inside at least one basic polygon. Since any point inside a basic polygon can be obtained by a policy, the point $C\in\mathcal{R}$. From the arbitrariness of $C$, we have $\mathcal{R}\supseteq\mathcal{C}$.

From Part II.1 and Part II.2, it can be proven that $\mathcal{R}=\mathcal{C}$. Since there are only finite deterministic policies in total, the set $\mathcal{R}$ is a convex polygon with its vertices all obtained by deterministic scheduling policies.

\textbf{Part III. Adjacent Vertices of $\mathcal{R}$}

For any two adjacent vertices $Z_{\boldsymbol{F}}$ and $Z_{\boldsymbol{F}'}$ of $\mathcal{R}$, if $\boldsymbol{F}$ and $\boldsymbol{F}'$ are different in more than one state, the polygon generated by them is convex. If the line segment $\overline{Z_{\boldsymbol{F}}Z_{\boldsymbol{F}'}}$ is inside the generated polygon, $Z_{\boldsymbol{F}}$ and $Z_{\boldsymbol{F}'}$ are impossible to be adjacent. If the line segment $\overline{Z_{\boldsymbol{F}}Z_{\boldsymbol{F}'}}$ is on the boundary of the generated polygon, there will be other vertices between them, then $Z_{\boldsymbol{F}}$ and $Z_{\boldsymbol{F}'}$ are still not adjacent. Therefore, we can conclude that policies $\boldsymbol{F}$ and $\boldsymbol{F}'$ are deterministic and different in only one state.

\subsection{Properties of set $\mathcal{L}$}
In this subsection, we will prove that the optimal delay-power tradeoff curve $\mathcal{L}$ is piecewise linear, decreasing, and convex. The vertices of the curve are obtained by deterministic scheduling policies. Moreover, the policies corresponding to adjacent vertices of $\mathcal{L}$ take different actions in only one state.

\begin{IEEEproof}
\textbf{Monotonicity:}

Since $\mathcal{L}=\{(P,D)\in\mathcal{R}|\forall(P',D')\in\mathcal{R},\text{ either }P'\ge P\text{ or }D'\ge D\}$, for any $(P_1,D_1),(P_2,D_2)\in\mathcal{L}$ where $P_1<P_2$, we should have $D_1\ge D_2$. Therefore $\mathcal{L}$ is decreasing.

\textbf{Convexity:}

Since $\mathcal{R}$ is a convex polygon, for any $(P_1,D_1),(P_2,D_2)\in\mathcal{L}$, their convex combination is $(\theta P_1+(1-\theta) P_2,\theta D_1+(1-\theta) D_2)\in\mathcal{R}$. Hence there exists a point $(P_\theta,D_\theta)$ on $\mathcal{L}$ where $P_\theta = \theta P_1+(1-\theta) P_2$, and $D_\theta\le \theta D_1+(1-\theta) D_2$. Therefore $\mathcal{L}$ is convex.

\textbf{Piecewise Linearity:}

Since $\mathcal{R}$ is a convex polygon, it can be expressed as the intersection of a finite number of halfspaces, i.e., $\mathcal{R}=\bigcap_{i=1}^{I}\{(P,D)|a_i P+b_i D \ge c_i\}$. We divide $(a_i,b_i,c_i)$ into 2 categories according to the value of $a_i$ and $b_i$ as $(a_i^+,b_i^+,c_i^+)$ for $i=1,\cdots,I^+$ if $a_i>0$ and $b_i>0$, and $(a_i^-,b_i^-,c_i^-)$ for $i=1,\cdots,I^-$ if $a_i\le 0$ or $b_i\le 0$. We have $I=I^++I^-$ and $I^+,I^->0$. Then $\mathcal{R}=\bigcap_{i=1}^{I^+}\{(P,D)|a_i^+ P+b_i^+ D \ge c_i^+\}\cap\bigcap_{i=1}^{I^-}\{(P,D)|a_i^- P+b_i^- D \ge c_i^-\}$. For $1\le l \le I^+$, define $\mathcal{L}_l=\{(P,D)|a_l^+ P+b_l^+ D = c_l^+\}\cap\bigcap_{i=1,i\neq l}^{I^+}\{(P,D)|a_i^+ P+b_i^+ D \ge c_i^+\}\cap\bigcap_{i=1}^{I^-}\{(P,D)|a_i^- P+b_i^- D \ge c_i^-\}$.

For all $(P,D)\in \mathcal{L}_l$, we have $(P,D)\in \mathcal{R}$. For all $(P',D')\in \mathcal{R}$, since $a_l^+ P'+b_l^+ D' \ge c_l^+=a_l^+ P+b_l^+ D$, it should hold that $P'\ge P$ or $D'\ge D$. According to the definition of $\mathcal{L}$, we have $(P,D)\in\mathcal{L}$. Therefore $\mathcal{L}_l \subseteq \mathcal{L}$.

For all $(P,D)\in \mathcal{L}$, we consider three cases: 1) If $a_i^+ P+b_i^+ D > c_i^+$ for all $1 \le i \le I^+$ and $a_i^- P+b_i^- D > c_i^-$ for all $b_i^->0$, set $\epsilon=\min_{b_i>0}\frac{a_i P+b_i D-c_i}{b_i}$ so that $a_i P+b_i (D-\epsilon) \ge c_i$ for all $b_i>0$. Since $(P,D)\in\mathcal{R}$, for all $b_i\le 0$ $a_{i} P+b_{i} D \ge c_i$, therefore $a_{i} P+b_{i} (D-\epsilon) \ge c_i$ for all $b_i\le 0$. Hence $(P,D-\epsilon)\in\mathcal{R}$, which is against the definition of $\mathcal{L}$. 2) If $a_i^+ P+b_i^+ D > c_i^+$ for all $1 \le i \le I^+$ and $a_i^- P+b_i^- D > c_i^-$ for all $a_i^->0$, set $\epsilon=\min_{a_i>0}\frac{a_i P+b_i D-c_i}{a_i}$ so that $a_i (P-\epsilon)+b_i D \ge c_i$ for all $a_i>0$. Since $(P,D)\in\mathcal{R}$, for all $a_i\le 0$ $a_{i} P+b_{i} D \ge c_i$, therefore $a_{i} (P-\epsilon)+b_{i} D \ge c_i$ for all $a_i\le 0$. Hence $(P-\epsilon,D)\in\mathcal{R}$, which is against the definition of $\mathcal{L}$. 3) If $a_i^+ P+b_i^+ D > c_i^+$ for all $1 \le i \le I^+$, and there exists $i^*$ and $j^*$ such that $a_{i^*}^-\le0$, $b_{i^*}^->0$, $a_{j^*}^->0$, $b_{j^*}^-\le0$, $a_{i^*}^- P+b_{i^*}^- D = c_{i^*}^-$, $a_{j^*}^- P+b_{j^*}^- D = c_{j^*}^-$. For all $(P',D')\in \mathcal{R}$, either $P'\ge P$, $D'\ge D$ or $P'\le P$, $D'\le D$. If there exists $P'<P$ and $D'<D$, then $(P,D)$ is against the definition of $\mathcal{L}$. If $P'\le P$ and $D'\le D$ for all $(P',D')$, since for all $1 \le i \le I^+$, we have $a_i^+ P+b_i^+ D > c_i^+$, therefore $a_i^+ P'+b_i^+ D' > c_i^+$. Hence $\mathcal{L}_i\cap\mathcal{R}=\emptyset$, which is against the condition. From the above three cases, for all $(P,D)\in \mathcal{L}$, there exists at least one certain $l^*$ such that $a_{l^*}^+ P+b_{l^*}^+ D = c_{l^*}^+$, which means $(P,D)\in\mathcal{L}_{l^*}$.

From above we can see that $\mathcal{L}=\bigcup_{l=1}^{I^+}\mathcal{L}_l$. Therefore $\mathcal{L}$ is piecewise linear.

\textbf{Properties of Vertices of $\mathcal{L}$:}

The vertices of $\mathcal{L}$ are also the vertices of $\mathcal{R}$, and adjacent vertices of $\mathcal{L}$ are also adjacent vertices of $\mathcal{R}$. From the results in Section B, vertices of $\mathcal{L}$ are obtained by deterministic scheduling policies, and the policies corresponding to adjacent vertices of $\mathcal{L}$ are different in only one state.
\end{IEEEproof}

\section{Proof of Theorem 2}
\label{appen_unconstrained_threshold}
\begin{IEEEproof}
From the literature, it is proven that there exists an optimal deterministic stationary policy. Therefore, in the proof, we only consider deterministic policies. Let $s(q)$ denote the transmitting packet number when $q[n]=q$. Define
\begin{align}
& h^{(m+1)}(q,s)\nonumber\\
=& q+ \mu P_s+\sum_{a=0}^A \alpha_a[h^{(m)}(q-s+a)-h^{(m)}(a)].
\end{align}

In the following, we will apply a nested induction method to prove the theorem, which utilizes the policy iteration algorithm for Markov Decision Processes. For a Markov Decision Process considering an average cost, the policy iteration algorithm, which is shown in Algorithm \ref{algo_policy_iteration}, can always converge to the optimal scheduling policy in finite steps, which is proven in \cite[Theorem 8.6.6]{puterman2014markov} and \cite[Proposition 3.4]{bertsekas1995dynamic}. In the algorithm, the function $h^{(m)}(q)$ will in the final converge to $h(q)$, which is normally known as the potential function or the bias function of the Markov Decision Process. The function $h(q)$ can be interpreted as the expected total difference between the cost starting from a specific state and the stationary cost.

We sketch the proof as follows. Initially, we assign $h^{(0)}(q)$ as a strictly convex function in $q$. Then in Part I, it will be demonstrated by the mathematical induction method that, in the policy improvement step of the policy iteration algorithm, for any $m$, if $h^{(m)}(q)$ is strictly convex in $q$, then $s^{(m+1)}(q)$ has the threshold-based property. On the other hand, in Part II, we show that in the policy evaluation step of the policy iteration algorithm, if $s^{(m+1)}(q)$ has the threshold-based property, then $h^{(m+1)}(q)$ is strictly convex in $q$. Based on the above derivations, we can prove the required conclusion by mathematical induction.

\begin{algorithm}[t]
\caption{Policy Iteration Algorithm for Markov Decision Processes}
\begin{algorithmic}[1]
\State $m \gets 0$
\ForAll{$q$}
\State $h^{(0)}(q) \gets$ arbitrary value // Initialization
\EndFor
\Repeat
\ForAll{$q$}
\State // Policy Improvement:
\State $s^{(m+1)}(q) \gets \arg\min_s\{h^{(m+1)}(q,s)\}$
\EndFor
\ForAll{$q$}
\State // Policy Evaluation:
\State $h^{(m+1)}(q) \gets h^{(m+1)}(q,s^{(m+1)}(q))$
\EndFor
\State $m \gets m+1$
\Until{$s^{(m)}(q)=s^{(m-1)}(q)$ holds for all $q$}
\State $s(q) \gets s^{(m)}(q)$ for all $q$
\end{algorithmic}
\label{algo_policy_iteration}
\end{algorithm}

\textbf{Part I. The Policy Improvement Step: Convexity of $h^{(m)}(q)$ in $q$ $\rightarrow$ threshold-based property of $s^{(m+1)}(q)$}

Assume $h^{(m)}(q)$ is strictly convex in $q$. In the following, we will show that $s^{(m+1)}(q)$ has the threshold-based property.
\begin{enumerate}
\item For a feasible policy, we have $s^{(m+1)}(0)=0$, and $s^{(m+1)}(1)=0$ or $1$. Therefore $s^{(m+1)}(q+1)-s^{(m+1)}(q)=0$ or $1$ when $q=0$.
\item Define $s_1=s^{(m+1)}(q_1)$ for a specific $q_1$. According to the Policy Improvement step, we have inequalities
\begin{align}
& h^{(m+1)}(q_1,s_1) \nonumber\\
\le & h^{(m+1)}(q_1,s_1-\delta), \forall 0\le \delta \le s_1,\label{eqn_t1_left}\\
& h^{(m+1)}(q_1,s_1) \nonumber\\
\le & h^{(m+1)}(q_1,s_1+\delta), \forall 0\le \delta \le S-s_1.\label{eqn_t1_right}
\end{align}

Since $h^{(m)}(q)$ is strictly convex in $q$, we have
\begin{align}
& h^{(m)}(q_1+1-s_1+a)-h^{(m)}(q_1-s_1+a)\nonumber\\
< & h^{(m)}(q_1+1-(s_1-\delta)+a)\nonumber\\
& -h^{(m)}(q_1-(s_1-\delta)+a), 0\le a\le A.
\label{eqn_h_convex}
\end{align}

Since $P_s$ is strictly convex, we have
\begin{align}
P_{s_1+1}-P_{s_1}<P_{s_1+1+\delta}-P_{s_1+\delta}.\label{eqn_Ps_convex}
\end{align}
From (\ref{eqn_t1_left}) and (\ref{eqn_h_convex}), we can obtain that
\begin{align}
& h^{(m+1)}(q_1+1,s_1)\nonumber\\
< & h^{(m+1)}(q_1+1,s_1-\delta), \forall 0 \le \delta \le s_1.\label{eqn_t1+1_left}
\end{align}

From (\ref{eqn_t1_right}) and (\ref{eqn_Ps_convex}), we can obtain that
\begin{align}
& h^{(m+1)}(q_1+1,s_1+1)\nonumber\\
< & h^{(m+1)}(q_1+1,s_1+1+\delta), \forall 0 \le \delta \le S-s_1-1.\label{eqn_t1+1_right}
\end{align}

From (\ref{eqn_t1+1_left}) and (\ref{eqn_t1+1_right}), it is shown that $s^{(m+1)}(q_1+1)$ can only be $s_1$ or $s_1+1$. That is to say, we have $s^{(m+1)}(q_1+1)-s^{(m+1)}(q_1)=0$ or $1$.
\end{enumerate}

From the above derivations, we prove by mathematical induction that $s^{(m+1)}(q)$ has the threshold-based property.

\textbf{Part II. The Policy Evaluation Step: Threshold-based property of $s^{(m+1)}(q)$ $\rightarrow$ convexity of $h^{(m+1)}(q)$ in $q$}

Assume $s^{(m+1)}(q)$ has the threshold-based property. We continue to use the same notation as in Part I, i.e., define $s_1=s^{(m+1)}(q_1)$ for a specific $q_1$, and $s^{(m+1)}(q_1+1)=s_1$ or $s_1+1$.
\begin{enumerate}
\item If $s^{(m+1)}(q_1+1)=s_1$,
\begin{align}
& h^{(m+1)}(q_1+1)-h^{(m+1)}(q_1)\nonumber\\
\le & h^{(m+1)}(q_1+1, s_1+1)-h^{(m+1)}(q_1,s_1)\\
= &  (q_1+1)+ \mu P_{s_1+1}\nonumber\\
& +\sum_{a=0}^A \alpha_a[h^{(m)}((q_1+1)-(s_1+1)+a)-h^{(m)}(a)]\nonumber\\
& -[ q_1+ \mu P_{s_1}+\sum_{a=0}^A \alpha_a[h^{(m)}(q_1-s_1+a)-h^{(m)}(a)]\\
= & 1+\mu (P_{s_1+1}-P_{s_1}).
\end{align}
On the other hand,
\begin{align}
& h^{(m+1)}(q_1+1)-h^{(m+1)}(q_1)\nonumber\\
>& h^{(m+1)}(q_1+1,s_1)-h^{(m+1)}(q_1,s_1-1)\\
=& (q_1+1)+\mu P_{s_1}\nonumber\\
& +\sum_{a=0}^A \alpha_a[h^{(m)}((q_1+1)-s_1+a)-h^{(m)}(a)]\nonumber\\
& -[q_1+ \mu P_{s_1-1}\nonumber\\
& +\sum_{a=0}^A \alpha_a[h^{(m)}(q_1-(s_1-1)+a)-h^{(m)}(a)]\\
=& 1+\mu (P_{s_1}-P_{s_1-1}).
\end{align}
\item If $s^{(m+1)}(q_1+1)=s_1+1$,
\begin{align}
& h^{(m+1)}(q_1+1)-h^{(m+1)}(q_1)\nonumber\\
=& (q_1+1)+ \mu P_{s_1+1}\nonumber\\
& +\sum_{a=0}^A \alpha_a[h^{(m)}((q_1+1)-(s_1+1)+a)-h^{(m)}(a)]\nonumber\\
& -[q_1+ \mu P_{s_1}+\sum_{a=0}^A \alpha_a[h^{(m)}(q_1-s_1+a)-h^{(m)}(a)]\\
=&1+ \mu (P_{s_1+1}-P_{s_1}).
\end{align}
\end{enumerate}

In conclusion, for any specific $q_1$, we have
\begin{align}
& 1+\mu(P_{s_1}-P_{s_1-1}) < h^{(m+1)}(q_1+1)-h^{(m+1)}(q_1) \le\nonumber\\
& 1+\mu(P_{s_1+1}-P_{s_1}).
\end{align}
That is to say, $h^{(m+1)}(q+1)-h^{(m+1)}(q)$ is strictly increasing. Therefore, we have $h^{(m+1)}(q)$ is strictly convex in $q$.

Based on our assumption for the initial value $h^{(0)}(q)$, as well as the derivations in Part I and II, it is proven by mathematical induction that $s^{(m)}(q)$ holds the threshold-based property for all $m \ge 1$. Since $s^{(m)}(q)$ converges to the optimal policy $s(q)$ for sure in finite steps, the optimal policy $s(q)$ will also hold the threshold-based property.
\end{IEEEproof}

\footnotesize
\bibliographystyle{IEEEtran}
\bibliography{IEEEabrv,scheduling}
\normalsize

\end{document}